\documentclass[11pt]{article}
\usepackage{graphics}
\usepackage{psfrag}
\usepackage{epsfig}
\usepackage{psfig}
\usepackage{epsf}
\usepackage{float}
\usepackage{rotating}
\newcommand{\beq}{\begin{equation}}
\newcommand{\eeq}{\end{equation}}
\newcommand{\beqa}{\begin{eqnarray}}
\newcommand{\eeqa}{\end{eqnarray}}

\newcommand{\vs}{\vspace{-0.2cm}}

\newcommand{\no}{\nonumber}
\newcommand{\Mp}{M_\pi}
\newcommand{\Mpn}{M_{\pi^0}}
\newcommand{\Mpp}{M_{\pi^+}}
\newcommand{\Mpz}{M_\pi^2}
\newcommand{\Mpnz}{M_{\pi^0}^2}
\newcommand{\Mppz}{M_{\pi^+}^2}

\newcommand{\LOI}{{\rm LO}\!\!\!\!/}

\begin{document}

\hfill FZJ-IKP(TH)-2000-21

\vspace{1in}

\begin{center}

{{\Large\bf Charge--dependent nucleon--nucleon potential\\[0.3em]
from chiral effective field theory}}

\end{center}

\vspace{.3in}

\begin{center}

{\large 
Markus Walzl,\footnote{email~ m.walzl@fz-juelich.de}
Ulf-G. Mei{\ss}ner,\footnote{email:~Ulf-G.Meissner@fz-juelich.de}
Evgeny Epelbaum\footnote{email:~evgeni.epelbaum@tp2.ruhr-uni-bochum.de}
}

\bigskip

{\it Forschungszentrum J\"ulich, Institut f\"ur Kernphysik 
(Theorie)\\ D-52425 J\"ulich, Germany} 

\end{center}

\vspace{.6in}

\thispagestyle{empty}

\begin{abstract}
\noindent 
We discuss charge symmetry and charge independence breaking in a 
chiral effective field theory approach for few--nucleon systems based
on a modified Weinberg power counting. We construct a
two--nucleon potential with bound and scattering states generated by
means of a properly regularized Lippmann--Schwinger equation.
We systematically introduce strong isospin--violating and electromagnetic 
operators in the theory. We use standard procedures to treat the
Coulomb potential between two protons in momentum space.
We present results for phase shifts in the
proton-proton, neutron-proton and the neutron-neutron systems.
We discuss the various contributions to
charge dependence  and charge symmetry breaking
observed in the nucleon--nucleon scattering lengths.
\end{abstract}

\vfill

\pagebreak

%%%%%%%%%%%%%%%%%%%%%%%%%%%%%%%%%%%%%%%%%%%%%%%%%%%%%%%%%%%%%%%%%%%%%%%%%%%%%%%%%
\section{Introduction}
\def\theequation{\arabic{section}.\arabic{equation}}
\setcounter{equation}{0}
\label{sec:intro}

The forces between two (or more) nucleons have been studied over many
decades but the connection to the underlying theory of the strong
interactions is so far only loosely established.  QCD is
formulated in terms of strongly interacting quark and gluon fields
and thus not amenable to most standard field theoretical
methods.\footnote{In principle, lattice gauge theory can be used to
study this problem but so far it has not attracted much attention. In
any case, a purely numerical solution to the problem always needs to
be supplemented by additional physical insight obtained using other
methods.} On the other hand, it is well established that
meson theory works fairly well at low energies. Such models are based
on the exchange of pions and heavier mesons between the nucleons
supplemented by form factors or short distance regulators. 
The drawback with such approaches is the fact that one can not
systematically calculate  corrections and that strong form factors
are ill-defined concepts in any field--theoretical sense. However, the pion
exchange originally proposed by Yukawa can now be linked to one
specific property of low--energy QCD, namely its spontaneous and
explicit chiral symmetry violation. Pions arise as Goldstone bosons
with point--like derivative couplings to the nucleons in accordance with
Goldstones theorem. In fact, one has a hierarchy of scales in the
problem, the long distances given by one-- and two--pion exchange,
whereas intermediate and short distances are often modelled in terms of
heavy meson exchanges and also are given by the unphysical form factor
ranges. Starting from nucleon and pion degrees of freedom,
effective field theory (EFT) allows  for a clean separation of these
scales and consequently offers a {\it systematic} and {\it controlled}
understanding of the forces between two or more nucleons.  Apart from
dealing with the various scales appearing in nuclear systems, it is straightforward
to implement the spontaneously and explicitely broken chiral symmetry of QCD 
as well as external probes in the EFT.
The appearance of shallow bound states (or, equivalently, large scattering
lengths) requires some method of resummation. There are essentially
two ways of tackling this problem.  One approach is
to build the potential from EFT and employ it in a properly
regularized Lippmann--Schwinger (or Schr\"odinger)
equation. This scheme was proposed in ref.\cite{wein} and will be referred to
as the Weinberg approach. Alternatively, one can also do the expansion directly on the
level of the scattering amplitude and resum the leading order, momentum--independent
four--nucleon interaction~\cite{KSW}. This is the so--called KSW scheme.
While at extremely low energies, much below the
scale set by the pion mass, it is sufficient to consider
four--nucleon interactions only (otherwise the rather successfull effective range
expansion would not work), for typical nuclear momenta of the size
of the pion mass, pions have to be included explicitely. In
refs.\cite{EGMI,EGMII} a next--to--next--to--leading order chiral
two--nucleon potential in a modified Weinberg scheme was 
developed  for the neutron--proton ($np$) system and shown to be close
in accuracy to the so--called modern potentials (in some partial waves). 
This isospin symmetric
potential will be the starting point of the investigation presented here.

\medskip\noindent
It is well established that the nucleon--nucleon interactions
are charge dependent (for a review, see e.g.\cite{jerry} and a recent
summary is given in~\cite{SAC}). For example, in 
the $^1S_0$ channel one has for the scattering lengths $a$ and the effective
ranges $r$
\beqa\label{CIBval}
a_{\rm CIB} &=& \frac{1}{2} \left( a_{nn} + a_{pp} \right) - a_{np}
= 5.64 \pm 0.40~{\rm fm}~,\\
r_{\rm CIB} &=& \frac{1}{2} \left( r_{nn} + r_{pp} \right) - r_{np}
= 0.03 \pm 0.06~{\rm fm}~.
\eeqa
These numbers for charge independence breaking ({\em CIB})
are taken from the recent compilation of Machleidt~\cite{CD-BonnII}.
It is understood that the Coulomb effects
for $pp$ ($nn$) scattering are subtracted based on standard methods (for a 
treatment of the Coulomb effects in proton--proton scattering in an
EFT framework,  see ref.\cite{RK}). One
sometimes refers to these as the nuclear amplitudes. 
The charge independence breaking in the scattering lengths is large, of
the order of 25\%, since $a_{np} = (-23.740 \pm 0.020)\,$fm. Of course,
it is magnified at threshold due to kinematic factors (as witnessed
by the disappearance of the effect in the effective
range)\footnote{Sometimes, CIB is simply discussed as the difference
  between the Coulomb subtracted $pp$ and the $np$ observables, which
  has to be accounted for when comparing to other values obtained in
  the literature. To avoid confusion, we will always explicitely state
  the definition we are using}. 
In addition, there are
charge symmetry breaking ({\em CSB}) effects leading to different values for
the $pp$ and $nn$ threshold parameters, 
\beq\label{CSBval}
a_{\rm CSB} =  a_{pp} - a_{nn} = 1.6 \pm 0.6~{\rm fm}~,\quad
r_{\rm CSB} =  r_{pp} - r_{nn} = 0.10 \pm 0.12~{\rm fm}~.
\eeq
Combining these numbers gives as central values $a_{nn} = -18.9\,$fm
and $a_{pp} = -17.3\,$fm. We should point out already here
that a recent measurement performed at Bonn~\cite{Witsch} leads to a sizeably
smaller value for $a_{nn}$ in the $^1S_0$ channel which affects
both values $a_{\rm CIB}$ and $a_{\rm CSB}$. We come back to
the implications of that measurement later.
Both the CIB and CSB effects have been 
studied intensively within potential
models of the nucleon--nucleon (NN) interactions. In such approaches, the
dominant CIB comes from the charged to neutral pion mass difference in the
one--pion exchange (OPE), $a_{\rm CIB}^{\rm OPE} \simeq 3.6 \pm 0.2\,$fm. 
Additional contributions come from $\gamma\pi$ and $2\pi$ (TPE) exchanges. 
According to some calculations, their size is approximately $1/3^{\rm rd}$ 
of the leading OPE contribution. For the TPE, significantly smaller results
can also be found in the literature, see e.g. table~3.3 in ref.\cite{jerry}.
In case of the $\pi\gamma$ contribution, a recent calculation\cite{CF} gives
a value which is more than a factor of ten smaller than  the 
leading OPE contribution.  We will come back to these issues later.
Note also that the charge dependence in the pion--nucleon coupling constants 
(if existing) in OPE and TPE
almost entirely cancel. Naively, one would expect a possible charge dependence
of the pion--nucleon coupling constants to be of a similar importance as the
pion mass difference in OPE. This strong suppression of
charge--dependent couplings has eluded a deeper understanding until the
EFT considerations based on power counting spelled out in ref.\cite{EM}. 
In the meson--exchange picture, CSB originates mostly from  
$\rho-\omega$ mixing, $a_{\rm CSB}^{\rho-\omega} \sim 1.2 \pm 0.4\,$fm. Other
contributions due to $\pi-\eta$, $\pi-\eta'$ mixing or the proton--neutron
mass difference are known to be much smaller. This picture has been
challenged in ref.\cite{LiMach} where it was claimed that the sum of various
two--boson--exchange diagrams can fully explain the observed CSB in the singlet
scattering length (for a more detailed discussion and a comparison to
the approaches based on $\rho-\omega$ mixing, see ref. \cite{CD-BonnII}).
It was already shown in 
refs.\cite{vanKolck,vKNij,FvK,EM} that EFT gives novel insight
concerning  the size of the TPE and $\pi\gamma$
contributions as well as the suppression of possible charge--dependent $\pi N$
coupling constants. However, a systematic and detailed study of 
CIB and CSB in the Weinberg 
approach has not been given so far. We will fill this gap by working out
the complete effects of isospin violation at leading order (LO) and
next--to--leading order (NLO) in the
modified Weinberg approach of refs.\cite{EGMI,EGMII}, which has been
proven very successful up to the pion production threshold, i.e. $E_{\rm lab}
\simeq 300\,$MeV.

\medskip\noindent
The manuscript is organized as follows. In section~\ref{sec:NCHPT} we
give a brief reminder of the effective field theory underlying our
calculations. The problem of isospin violation in the NN force is
taken up in section~\ref{sec:isov}. We show how to incorporate strong
and electromagnetic isospin breaking in the EFT based on an extended
power counting and organize the various contributions from the light
quark mass difference as well as from soft and hard photons. In
addition, we discuss how to treat the Coulomb potential in momentum
space. In section~\ref{sec:pot} the renormalized potential at NLO is
displayed. It consists of one-- and two--pion exchanges, $\pi \gamma$
graphs and local four--nucleon interactions with zero or two
derivatives. The results are given and discussed in
section~\ref{sec:res}. We conclude with a summary and outlook.
Some subtleties related to the CIB two--pion exchange are spelled out in the
appendix.

\section{Brief reminder of nuclear chiral perturbation theory}
\def\theequation{\arabic{section}.\arabic{equation}}
\setcounter{equation}{0}
\label{sec:NCHPT}

In this section we briefly spell out the central ideas underlying our
calculations.  One starts from an effective chiral Lagrangian  of 
pions and nucleons, including in particular local
four--nucleon interactions which describe the short range part of the nuclear 
force, symbolically
\beq
{\cal L}_{\rm eff}  = {\cal L}_{\pi\pi} + {\cal L}_{\pi N} + {\cal L}_{NN}~,
\eeq
where each of the terms admits an expansion in small momenta and quark (meson)
masses. To a given order, one has to include all terms consistent with chiral
symmetry, parity, charge conjugation and so on.
The methods to construct such Lagrangians and to calculate 
the corresponding Feynman diagrams are by now standard, see
e.g.~\cite{bkmrev}. From the effective Lagrangian, one
derives the two--nucleon potential.
This is based on a modified Weinberg counting as described
in ref.\cite{EGMI} which is applied to the  two--nucleon potential
to a certain order in small momenta and pion masses,
\beq 
V(\vec{p},\vec{p}\,') = \sum_i V^{(i)} (\vec{p},\vec{p}\,')~,
\eeq
with $\vec{p},\vec{p}\,'$ the nucleon centre-of-mass (cms) momenta 
and the superscript
$i$ gives the (non--negative) chiral dimension. The power counting
underlying this potential is based on the considerations presented in
ref.\cite{EGMI}. To leading order (LO), this
potential is the sum of the time--honored one--pion exchange (OPE)
(with  point-like coupling)  and of
two four--nucleon contact interactions without derivatives. The
low--energy constants (LECs) accompanying these terms have to be
determined by a fit to some data, like e.g. the two S-waves in the
low energy region (for $np$). At
next--to--leading order (NLO), one has corrections to the OPE, the leading
order two--pion exchange (TPE) graphs (called the box, triangle and football
diagrams according to the number of non--linear $NN\pi\pi$ interactions
involved, which are zero, one and two, in order) and seven dimension
two four--nucleon terms with unknown LECs (for the $np$ system).  
The existence of shallow nuclear bound states (or large scattering 
lengths $a \gg 1/M_\pi$) 
forces one to perform an additional nonperturbative resummation. This is done
here by obtaining the bound and scattering states by solving a regularized 
Lippmann--Schwinger (LS) equation, projected onto states with orbital
angular momentum $l$, total spin $s$ and total angular momentum $j$, 
\beq\label{LSeq}
T^{sj}_{l',l} (p',p) = V^{sj}_{l',l} (p',p) + \sum_{l''} \,
\int_0^\infty \frac{dp'' \, {p''}^2}{(2 \pi )^3} \,  V^{sj}_{l',l''} (p',p'')
\frac{m}{p^2-{p''}^2 +i\eta} T^{sj}_{l'',l} (p'',p)~,
\eeq
with $\eta \to 0^+$ and $m$ denotes the nucleon mass. 
In the uncoupled case, $l$ is conserved. The partial wave projected
potential $V^{sj}_{l',l} (p',p)$ is obtained by standard methods, see e.g.~\cite{EGMII}.
The potential has to be understood as regularized, as dictated
by the EFT approach employed here. That is done in the following way:
\beq\label{Vreg}
V( {p}, {p}'\,) \to f_R ( {p} ) \, V( {p}, {p}') \, f_R ({p}' )~,
%V( \vec{p},\vec{p}~'\,) \to f_R ( \vec{p}\,) \, V( \vec{p},\vec{p}~'\,) \,
%f_R (\vec{p}~'\,)~,
\eeq
where $f_R (p)$ is a regulator function chosen in harmony with the
underlying symmetries. In what follows, we work with a sharp cut--off,
\beq\label{reg1}
 f_R^{\rm sharp} (p) = \theta (\Lambda^2 -p^2)~.
\eeq
Within a certain range of cut--off values, the physics should be
independent of its precise value. This range increases has one goes to
higher orders, as demonstrated explicitely for the $np$ case in~\cite{EGMII}.

\section{Explicit isospin violation}
\def\theequation{\arabic{section}.\arabic{equation}}
\setcounter{equation}{0}
\label{sec:isov}

\subsection{Sources of isospin violation}

Here, we briefly discuss how isospin violation arises in the Standard Model and
show how it can be implemented in the effective field theory. We stress that there
are two small parameters which allows one to treat strong and electromagnetic
isospin breaking perturbatively. The electromagnetic effects can further be separated
into two types, soft and hard photons. The soft photons lead on one hand to
perturbatively calculable photon loop effects and on the other hand the
nonperturbative resummation of such photons generates the long range Coulomb
force. The effects of hard photons
can completely be absorbed in electromagnetic short distance  operators. 
We also briefly
discuss the relation to the notions of charge independence and charge symmetry 
more often used in this context.

\medskip \noindent
Consider first pure QCD, i.e. a world with no electroweak 
interactions.\footnote{We consider the quark masses as parameters and 
are not concerned with their dynamical origin here.} To be specific,
we are only concerned with the sector of the two light up and down quarks. For equal
current quark masses $m_u = m_d$, the QCD Hamiltonian is invariant under   
a global (flavor) transformation  of the type
\beq
q  = 
\left( \begin{array}{c} 
                 u \\ d

\end{array} \right) 
\rightarrow q' = {\cal V}\, q = {\cal V} \,
\left( \begin{array}{c} 
                 u \\ d
\end{array} \right) \,\,\, ,
\quad {\cal V} \in SU(2)
\quad .
\label{ud}
\eeq
Of course, in nature the light quark masses are not equal. In fact, they
differ considerably, $m_u/m_d = 0.55\pm 0.15$ for the standard 
$\overline{\rm MS}$ subtraction scheme
at a renormalization scale of 1~GeV. Still, isospin is a good approximate
symmetry because what counts is not the ratio of the quark masses but rather
their difference compared to the typical scale of strong interactions,
\beq 
\varepsilon \equiv {m_d-m_u \over \Lambda_{\rm QCD}} \simeq {1 \over 30} \ll 1~,
\eeq
since $\Lambda_{\rm QCD} \simeq 150\,$MeV. 
QCD thus provides a reason why isospin is such a good approximate symmetry. 
In fact, one could also use the larger chiral symmetry breaking scale
$\Lambda_\chi = 4\pi F_\pi / \sqrt{N_f}$, with $F_\pi = 92.4\,$MeV the pion
decay constant and $N_f$ the number of quark
flavors. This would indicate even smaller isospin breaking effects.
For our purposes it suffices to state that such dimensional arguments
let one understand the smallness of the observed isospin breaking.
Another way of looking at the relative strength of isospin violation is
to compare the isoscalar (isospin conserving) $\sim (m_u+m_d)$
to the isovector (isospin violating) $\sim (m_u-m_d)$
part of the QCD quark mass term,
\begin{equation}
{\cal H}_{\rm mass}^{\rm QCD} = -\frac{1}{2}\bar{q} \, 
(m_{\rm u}+m_{\rm d})(1+\epsilon\tau_{3})\,q~,
\end{equation}
leading one to expect  that 
\beq\label{epsdef}
\epsilon \equiv {m_d-m_u \over m_d+m_u} \sim {1 \over 3}
\eeq 
would be the appropriate small parameter. From the
numerical value of $\epsilon$, isospin symmetry would appear to be accidental. This 
argument, however, ignores the generic scale of QCD generated by dimensional
transmutation. On the other hand, for the organisation of the effective field theory,
it is advantageous to work in the isospin basis as given in eq.(\ref{epsdef}). Therefore,
we will count strong isospin violation in terms of  $\epsilon$, but keep in mind that
the true size of isospin violation is smaller than indicated by the numerical size
of the parameter $\epsilon$.

\medskip\noindent
In the presence of electromagnetism, further isospin
violation is induced since the charges of the quarks are unequal. These effects are
in general small due to the explicit appearance of the fine structure
constant $\alpha = e^2/4\pi = 1/137.036$. In fact, in many cases the strong and
electromagnetic isospin--violating effects are of the same size. A good
example is the neutron--proton mass difference where $(m_n-m_p)^{\rm QCD} \simeq 2.1\,$MeV
and $(m_n-m_p)^{\rm QED} \simeq -0.8\,$MeV, leading to the physical value of
$m_n - m_p = 1.3\,$MeV.  The pion mass difference is
special, because of the absence of $D$--couplings in SU(2) there is no strong
contribution in the light quark mass difference but only a very tiny second
order contribution (due to $\pi^0-\eta$ mixing). In what follows, we will use
\beq
\Delta M_\pi = M_{\pi^\pm} - M_{\pi^0} = 4.6~{\rm MeV}~,
\eeq
which is almost entirely of electromagnetic origin. Due to the long range nature
of the electromagnetic interactions, one has to be careful in separating long and
short distance physics generated by virtual photons.\footnote{By virtual photons we
refer to photon exchanges within any given Feynman diagram.} While the latter can be
represented by a string of local contact interactions with increasing chiral dimension,
the former generate the Coulomb singularities and need to be treated separately.

\medskip\noindent
In the nuclear physics language, it is more common to talk about charge (in)dependence
and charge symmetry (breaking). Charge independence refers to the invariance
under any rotation
in isospin space. A violation of this symmetry is called charge dependence or charge
independence breaking (CIB). On the quark level, charge 
symmetry is a rotation about the 2--axis in isospin space
\beq
{\cal P}_{\rm cs}
\left( \begin{array}{c} 
                 u \\ d
\end{array} \right) 
= {\rm e}^{(i\pi /2)\tau_2} \,
\left( \begin{array}{c} 
                 u \\ d
\end{array} \right) 
=
\left( \begin{array}{c} 
                 -d \\ \,\,u
\end{array} \right) \,\,\, ,
\quad {\cal P}_{\rm cs}^2 = {\bf 1}
\quad ,
\label{Pcs}
\eeq
and similarly for the nucleon isodoublet consisting of the proton and the
neutron. The violation
of this symmetry is called charge symmetry breaking (CSB). It is a special 
case of charge dependence. On the level of the two--nucleon force,
charge independence implies a potential of the form (i.e. which is a scalar 
in isospin space)
\beq
V_{ij} = A + B \vec{\tau}_i \cdot \vec{\tau}_j~,
\eeq
where $A$ and $B$ are functions of the nucleon spin and momentum (or space)
coordinates and `$i,j$' labels the interacting nucleons.  
Charge symmetry allows for a more general two--body force,
\beq
V_{ij, {\rm cs}} = A + B \vec{\tau}_i \cdot \vec{\tau}_j + C \tau_i^3
\cdot \tau_j^3~.
\eeq
If one works out the Coulomb potential in terms of the nucleon charge
matrix $Q$, it is obvious that the electromagnetic effects lead
to breaking of charge independence and charge symmetry,
\beq
V_{ij, {\rm Coulomb}} = {e^2\over 16 \pi r^2} (1+ \tau_i^3)
(1+ \tau_{j}^3)~.
\eeq 
Stated differently, CIB of the strong NN interactions means that the proton--proton ($pp$),
neutron--proton ($np$) or neutron--neutron ($nn$) are different once the Coulomb effects
have been removed. CSB, on the other hand, refers to the difference between the $pp$
and the $nn$ interactions.

\subsection{Power counting, effective Lagrangian and classification scheme}
\label{sec:class}
As discussed in the preceeding section, 
the Standard Model supplies us with two small parameters,
which allow one to systematically include isospin violation in terms 
of two external scalar sources,
\beqa\label{scalsource}
\chi &=& 2B \, (m_{\rm u}+m_{\rm d})\, (1+\epsilon\tau_{3})~, \\
Q &=& \frac{e}{2} \, (1 + \tau_3)~,\label{ss2}
\eeqa
where $B$ measures the strength of the quark-antiquark condensation in the
vacuum, $B = |\langle 0 | \bar{u}u|0\rangle |/F_\pi^2$, with $F_\pi$ the 
weak pion decay constant. We work here in the standard approach with 
$B \gg F_\pi$. The charge matrix $Q$ defines the strength
of the virtual photon coupled to the matter fields. Since a virtual photon can not leave
a Feynman diagram, only even powers of $Q$ can appear in the effective
Lagrangian (with the exception of the covariant derivative), as
detailed below. A further consequence is that virtual photons appear necessarily  
in loop diagrams. The germane electromagnetic expansion parameters is $\alpha =
e^2/4\pi$ (although one could equally well use $e^2$ or $e^2/(4\pi )^2$).

\medskip \noindent
Due to its perturbative nature induced by the small parameters, we
treat the strong and electromagnetic
isospin violation in addition to the power counting of the isospin symmetric potential 
mentioned in section~\ref{sec:NCHPT}. 
In principle, one has a double expansion of the form $\epsilon^n \, \alpha^m$ with 
$n,m$ are integers. However, since we iterate the potential in the LS equation,
we have to avoid double counting. Here, the isospin basis given in 
Eqs.(\ref{scalsource},\ref{ss2}) proves to be very useful.

\medskip \noindent
We now discuss the various parts of the effective
Lagrangian underlying the analysis of isospin violation in the two--nucleon
system. To include virtual photons in the pion and the pion--nucleon system  using the above-defined 
external sources  is by now a standard procedure~\cite{egpdr}-\cite{mm}.
The lowest order (dimension two)  pion Lagrangian
takes the form
\beq
{\cal L}_{\pi\pi} = \frac{F_\pi^2}{4} \langle \nabla_\mu U \nabla^\mu U^\dagger
+ \chi U^\dagger + \chi^\dagger U \rangle + C \langle QUQU^\dagger\rangle~.
\eeq 
Here, $U(x)$ collects the isotriplet of pion fields,  $\nabla_\mu$ is the (pion) covariant
derivative containing the virtual photons, $\langle \,\, \rangle$ denotes
the trace in flavor space,  and the last term, which contains the
nucleon charge matrix $Q$=$e\,$diag$(1,0)$,\footnote{Or equivalently, one can use the
quark charge matrix $e(1+\tau_3)/2$, see refs.\cite{ms,mm}.} 
leads to the charged to neutral
pion mass difference, $\Delta M_\pi^2 = M_{\pi^\pm}^2 - M_{\pi^0}^2$, via
$\Delta M_\pi^2 = 8\pi\alpha C/F_\pi^2$, i.e. $C = 5.9\cdot 10^{-5}\,$GeV$^4$. Note that
to this order the quark mass difference $m_u - m_d$ does not appear in the
meson Lagrangian (due to G--parity). That is chiefly the reason why the
pion mass difference is almost entirely an electromagnetic (em) effect.
The equivalent pion--nucleon Lagrangian to second order takes the form
\beqa
{\cal L}_{\pi N}^{\rm str} &=& N^\dagger \left( iD_0 - \frac{g_A}{2} 
\vec{\sigma}\cdot\vec{u} \right) N +
 N^\dagger \left\{ \frac{\vec{D}^2}{2m} + c_1 \langle \chi_+ \rangle
+\left( c_2 -\frac{g_A^2}{8m}\right) u_0^2 + c_3 u_\mu u^\mu
\right. \nonumber\\
&& \qquad \qquad + \left.\frac{1}{4} 
\left(c_4 + \frac{1}{4m}\right) [\sigma_i , \sigma_j ] u_i u_j
+ c_5 \left( \chi_+ - \frac{1}{2} \langle \chi_+ \rangle\right) + \dots
\right\}  N~,
\eeqa
which is the standard heavy baryon effective Lagrangian in the rest--frame
$v_\mu =(1,0,0,0)$. $m$ is the nucleon mass, $u = \sqrt{U}$,  $u_\mu$ the chiral
viel--bein, $u_\mu \sim -i\partial_\mu\phi/F_\pi + \ldots\,$ and $\chi_+ =
u \chi^\dagger u + u^\dagger \chi u^\dagger$. 
The four--nucleon interactions to be discussed below
do not modify the form of this Lagrangian (for a general discussion, see
e.g. ref.\cite{FMS}). Strong isospin breaking is due to the operator $\sim c_5$.
More precisely, this low--energy constant always appears in the
combination $c_5 \, B \, (m_u-m_d)$.
Electromagnetic terms to second order are given by~\cite{ms,mm}
\beq\label{LpiN2em}
{\cal L}_{\pi N}^{\rm em} = F_\pi^2 N^\dagger \left\{ f_1 \langle Q_+^2 
- Q_-^2\rangle + f_2 \tilde{Q}_+\langle Q_+\rangle + f_3 \langle Q_+^2
+ Q_-^2\rangle  \right\} N~,
\eeq 
with $Q_\pm = uQ^\dagger u \pm u^\dagger Q u^\dagger$ and $\tilde{A} = A -
\langle A\rangle /2$ projects onto the off--diagonal elements of the
operator $A$.  The last term
in eq.(\ref{LpiN2em}) is not observable since it leads to an equal
electromagnetic mass
shift for the proton and the neutron, whereas the operator $\sim f_2$ to this 
order gives the em proton--neutron mass difference. In what follows, we will
refrain from writing down such types of operators which only lead
to an overall shift of masses or coupling constants. We note that in the pion
and pion--nucleon sector, one can effectively count the electric charge as
a small momentum or meson mass. This is based on the observation that
$M_\pi / \Lambda_\chi \sim e/\sqrt{4\pi} = \sqrt{\alpha} \sim 1/10$.
It is thus possible to turn the dual
expansion in small momenta/meson masses on one side and in the electric coupling
$e$ on the other side into an expansion with one generic small parameter.
As noted before, we use the fine structure constant $\alpha
=e^2/4\pi$ as the electromagnetic expansion parameter. 
We now turn to the two--nucleon sector, i.e. the four--fermion contact
interactions without pion fields. Consider first the strong terms. 
Up to one derivative, the effective Lagrangian takes the form 
\beqa
{\cal L}_{NN}^{\rm str}&=& l_1 (N^\dagger N)^2 + l_2 (N^\dagger \vec{\sigma} N)^2
+ l_3 ( N^\dagger \langle \chi_+ \rangle N) (N^\dagger N) + l_4
(N^\dagger \tilde{\chi}_+ N)(N^\dagger N)
\nonumber \\
&+& l_5 ( N^\dagger \vec{\sigma}\langle \chi_+ \rangle N) (N^\dagger \vec{\sigma}
 N) + l_6 (N^\dagger \vec{\sigma} \tilde{\chi}_+ N)(N^\dagger \vec{\sigma} N)
+ \ldots~,
\eeqa
where the ellipsis denotes terms with two (or more) derivatives acting on the
nucleon fields. Similarly, one can construct the em terms. The ones without
derivatives on the nucleon fields  read
\beqa
{\cal L}_{NN}^{\rm em}&=& N^\dagger \left\{ r_1 \langle Q_+^2 - Q_-^2\rangle
+ r_2 \tilde{Q}_+ \langle Q_+\rangle \right\} N (N^\dagger N) 
\nonumber \\
&+& N^\dagger \vec{\sigma} \left\{ r_3 \langle Q_+^2 - Q_-^2\rangle 
+ r_4  \tilde{Q}_+ \langle Q_+\rangle \right\} N (N^\dagger \vec{\sigma} N)
\nonumber \\
&+&   N^\dagger  \left\{ r_{5}Q_+ + r_{6}\langle Q_+ \rangle \right\} 
N (N^\dagger Q_+ N) +
   N^\dagger \vec{\sigma} \left\{ r_{7}Q_+ + r_{8}\langle Q_+ \rangle 
\right\} N (N^\dagger \vec{\sigma} Q_+ N) \nonumber \\
&+& r_{9} (N^\dagger  Q_+ N)^2 + r_{10} (N^\dagger \vec{\sigma} Q_+ N)^2~.
\eeqa
There are also various terms resulting form the insertion of
the Pauli isospin matrices $\vec{\tau}$ in different ${N}^\dagger N$
binomials. Some of these can be eliminated by Fierz reordering, while
the others are of no importance for our considerations.
The special form  of eqs.(\ref{scalsource},\ref{ss2}) allows us to
rewrite all nucleonic contact terms with insertions of external fields at a given chiral 
order as the sum of isospin symmetric contact structures and of
\beqa
V_{\rm CSB} &\sim& (N^{\dagger}\tau_{3}N)(N^{\dagger}N),\\
V_{\rm CIB} &\sim& (N^{\dagger}\tau_{3}N)^{2},
\eeqa
which parameterize the non--pionic CSB and CIB-effects (as discussed in
more detail below).  Note that in the iteration all
isoscalar pieces in the external sources $Q$ and $\chi$ get resummed and we therefore
only consider terms linear in $\alpha$ and $\epsilon$. 
As in ref.\cite{EGMII}, we form appropriate combinations of these
four--nucleon operators so that we can pin down the corresponding LECs
in the low partial waves. Specifically, we combine the isospin
conserving and violating terms in the following manner (in a somewhat symbolic
notation),
\beq
{\cal L}_{NN} = C_\beta^{\rm sym} \, (N^\dagger \, P_\beta \, N)^2
 +  C_\beta^{\rm str} \, (N^\dagger \, P_\beta \, \langle
 \chi_+ \rangle \, N) \, (N^\dagger N)
 +  C_\beta^{\rm em} \, (N^\dagger \, P_\beta \, \langle
 Q^2 \rangle \, N) \, (N^\dagger N) + \ldots
\eeq
where the ellipsis stands for terms with two (or more) derivatives.
Also, $P_\beta$ is a projector onto the appropriate partial wave with
$\beta$ denoting the corresponding angular momentum and isospin
quantum numbers, i.e. for $\beta = \, ^1S_0$ we have $P_\beta = \sigma_2 
\tau_2 \vec{\tau} / \sqrt{8}$. The precise number of parameters in each
partial wave to a given order will be enumerated in the next section.
It is also worth to stress that the  isospin--breaking contact interactions 
appear at the same order as the four--nucleon contact terms with two
derivatives in the isospin symmetric case. For that, we have to specify
more explicitely how to organize the various isospin breaking contributions 
to the  potential.

\medskip\noindent
After having set up the power counting of  the underlying effective
Lagrangian, we are in the position of classifying the various contributions to
isospin violation (or CIB and CSB) in the NN interaction, more precisely,
on the level of the two--nucleon potential. The pion mass
difference in the OPEP leads to a contribution of the form
\beq
V_{\rm OPE}^{\Delta M_\pi} 
\sim \Delta M_\pi^2 {\partial \over \partial M_\pi^2} V_{\rm OPE}
\sim {\Delta M_\pi^2 \over  M_\pi^2} V_{\rm OPE} \sim \alpha Q^{-2}~,
\eeq 
where $Q$ denotes some small momentum or pion mass and $V_{\rm OPE}$
the isospin symmetric OPEP.  Therefore, the pion mass squared in 
the denominator leads to an enhancement of this
contribution by two powers in small momenta. So the leading order
isospin violating potential scales as $\alpha/Q^2$.
This order we
denote by  ``$\LOI$'', meaning leading order in the isospin breaking
NN potential. It is also worth to point out that we count the 
Coulomb potential as ${\rm LO}\!\!\!\!/$~ despite the
fact that it stems from an infinite sum of photon exchanges.  Formally,
however,  the Coulomb potential $\alpha / |\vec{q} - \vec{q}\,'|^2$ has
the same structure as the $\LOI$~ OPEP. The leading order momentum--independent
isospin violating four--nucleon contact interactions do not appear at 
$\LOI$~ because they simply scale as $\alpha Q^0$ and $\epsilon Q^0$, so they
come together with the pion mass difference and $\pi\gamma$ graphs at 
N$\LOI$~, since these latter two also scale as $\alpha Q^0$. Subleading TPE
proportional to the dimension two LECs $c_i$ would appear at N$^2\LOI$~ whereas
a possible isospin breaking in the pion--nucleon coupling constants,
$g_{\pi^0 pp} \neq g_{\pi^+ pn}$, is a N$^3\LOI$~ effect, i.e. is expected to 
be very suppressed.

\begin{center}
\begin{tabular}{|l|c|l|}\hline
Order & Parameter & Contribution \\ \hline
$\LOI$ & $\alpha$ & Pion mass difference in OPE 
\\ & $\alpha$ & Coulomb potential
\\ \hline
N$\LOI$ & $\alpha$ & Pion mass difference in TPE \\ 
& $\alpha$ &  $\pi \gamma$--exchange  \\ 
 & $\alpha$ & four--nucleon contact interaction with no derivatives $\sim 
(N^\dagger \tau_3  N)^2$ 
\\ & $\epsilon$ & four--nucleon contact interaction with no derivatives
$\sim (N^\dagger \tau_3 N)(N^\dagger N)$\\ 
\hline
\end{tabular}
\end{center}

\medskip\noindent
We remark that this power counting is similar to but also distinctively different
from the one in the KSW scheme~\cite{EM}. In that approach, pion exchange is treated
perturbatively and consequently $\pi\gamma$ loop graphs only appear at NN$\LOI$. 
It is instructive to  consider the N$\LOI$~ four--nucleon contact 
interactions (which only affect the S--waves). Terms
proportional to  $(N^\dagger Q N)^2$ or
$(N^\dagger \chi_+ N)^2$ clearly lead to CIB since
\beq
\langle nn |(N^\dagger \tau_3 N)^2 | nn \rangle =
\langle pp |(N^\dagger \tau_3 N)^2 | pp \rangle \neq 
\langle np |(N^\dagger \tau_3 N)^2 | np \rangle~,
\eeq
whereas the operator proportional to the light quark mass difference, i.e. the
isovector part of $\chi$ leads to CSB,
\beq
\langle nn |(N^\dagger \tau_3 N)(N^\dagger N) | nn \rangle =
-\langle pp |(N^\dagger \tau_3 N)(N^\dagger N) | pp \rangle \quad {\rm and} \quad
\langle np | (N^\dagger \tau_3 N)(N^\dagger N) | np \rangle = 0~.
\eeq
This means that the leading order mechanisms giving rise to CIB and CSB
are very different, in particular, CSB is driven by the light quark
mass difference. We should also note that in the above classification
scheme we have not listed the nucleon mass difference
explicitely. Throughout, the kinematics is given in terms of the
physical  proton and neutron mass, $m_p = 938.27\,$MeV and $m_n =
939.57\,$MeV, respectively, leading to slightly different threshold
energies. This is an effect concerning the external legs and therefore
affects  the S--matrix of the scattering process only
indirectly. Consequently, we did
not list such type of contribution in the classification scheme.
In the pertinent TPE  graphs we neglect this
mass splitting, which is consistent with the underlying power counting.
We also point out that the possible isospin breaking in the
pion--nucleon coupling constants appears at NN$\LOI$ in the KSW
scheme as first observed in ref.\cite{EM}. Before we can work out the effects
of CIB and CSB in the two--nucleon system, we have to discuss how to
treat the long--range Coulomb force.

\subsection{Momentum space treatment of the Coulomb potential}
\label{sec:VP}
As the Coulomb potential is of infinite range, the S--matrix has 
to be formulated in terms of asymptotic Coulomb states. Therefore, the
phase shifts for a given angular momentum $l$
due to the strong potential in the presence of the
long--range electromagnetic interactions, denoted by $\delta_l^l$,
are defined in terms of a linear combination of (ir)regular 
Coulomb--functions $F(G)$ as
\beq\label{ucwave}
\chi_{l} (r) = F_{l}(r) + \tan(\delta_l^l) \, G_{l}(r)
\eeq
analogously to the expression for an arbitrary potential of short range 
(i.e. in the absence of the Coulomb force)
\beq\label{plainasym}
\chi_{l} (r) = F^0_{l}(r) + \tan(\delta_l^s) \, G^0_{l}(r)
\eeq
with $F^0$, $G^0$ denoting solutions of the Coulomb problem with zero
charge (conventionally expressed in terms of Bessel and Neumann
functions) and the corresponding phase shift is called $\delta_l^s$.
So far, we have restricted ourselves to uncoupled channels, but nevertheless it 
is straightforward to extend the formalism to the case of coupled
ones. One simply has to rewrite the wave--functions as 2x2--matrices
\beq
\Psi\,\equiv\,\Psi_{l-1,l+1}\,=\,\left( \begin{tabular}{c c} 
                 $\Psi_{l-1}$ & 0 \\ 0 & $\Psi_{l+1}$
\end{tabular} \right) ~,
\eeq
with $\Psi =F, G$, substitute tan$\delta_l$ by the coupled-channel 
K--matrix $K_{l-1,l+1}$, and  redefine eq.(\ref{ucwave}) as a matrix-equation
(for compactness, we drop all indices)
\beq
\chi\,=\,F-\mu  \, q \,  K \, G~,
\eeq
with $\mu$ being the reduced mass and $q$ the momentum transfer.
In what follows we will only consider uncoupled channels  to simplify
notation, nevertheless the extension to coupled channels will
always be obvious in the above-described sense.\\

\noindent As eq.(\ref{ucwave}) exhibits  asymptotical Coulomb-states, 
we have to re-express our Lippmann-Schwinger-equation
in terms of them. Therefore, the calculation of phase shifts requires  a large number of 
Fourier-transformed Coulomb-functions. That would make the
investigations very time-consuming.
Consequently,  we are now going to develop a formalism that makes use
of the phase shifts calculated for plane wave asymptotics 
as defined in eq.(\ref{plainasym}). This procedure has been applied
first in ref.\cite{VP}.
The starting point of this technique is the observation, that for a potential of the form
\begin{equation}\label{VsC}
V = V_{\rm Coulomb} + V_{\rm strong}
\end{equation}
with 
\begin{equation}\label{cond}
V_{\rm strong}\psi_{l} = 0 \,\,\,(r\ge R)~,
\end{equation}
and $\psi_{l}$ the two--nucleon wave-function for a given angular momentum, 
two exact solutions for the wave-function can be given for every point
on a sphere with radius $R+\epsilon$. One is of the form as in eq.(\ref{ucwave}),
and another one according to eq.(\ref{plainasym}) with 
the phase shifts calculated for the following potential as in eq.(\ref{VsC}),
with V$_{\rm Coulomb}$ being the Fourier-transformed Coulomb-potential
integrated to the radius $R$, 
\beqa\label{Coulombpot}
V_{\rm Coulomb}(\mid \vec{q}\,'- \vec{q}\mid) &=&  
\int_{0}^{R}d^{3}r \,  e^{i(\vec{q}\,'-\vec{q}\,) \cdot
\vec{r}}\,\,\frac{\alpha}{r} \nonumber\\
                               &=& 
\frac{4\pi \alpha}{\mid \vec{q}~'-\vec{q}\mid^{2}}
(1-\cos(\mid \vec{q}~'-\vec{q}\mid R))~.
\eeqa
Here, $\vec{q}, \vec{q}\,'$ are the cms momenta and $\alpha$ is the fine-structure
constant. On the above-defined sphere, both wave functions describe
the same system. 
Now we know how to obtain an expression for the strong shift
$\delta_l^l$ in the presence of the Coulomb interaction
in terms of the short-range shift $\delta_l^s$ in the absence of
electromagnetism: We only have to match the two solutions. 
This is most conveniently done by requiring the logarithmic derivative
of both solutions to be equal, what enables us to express the strong
shift in the presence of the Coulomb force in a Wronskian form:
\beq
\tan(\delta_l^l)\,=\,\frac{\tan(\delta_l^s)\lbrack 
F,G_0\rbrack + \lbrack F,F_0\rbrack}{\lbrack F_0,G\rbrack 
+ \tan(\delta_l^s)\lbrack G_0,G\rbrack}
\eeq
with
\beq
\lbrack F,G \rbrack\,=\,\biggl(G\frac{dF}{dr}-F\frac{dG}{dr}\biggr)_{r=R}
\eeq
Here again, the continuation to coupled channels in the form of a
matrix-equation is obvious. 

\begin{figure}[htb] 
\begin{center}
%\hspace{-1cm}
\begin{turn}{270}
\epsfig{file=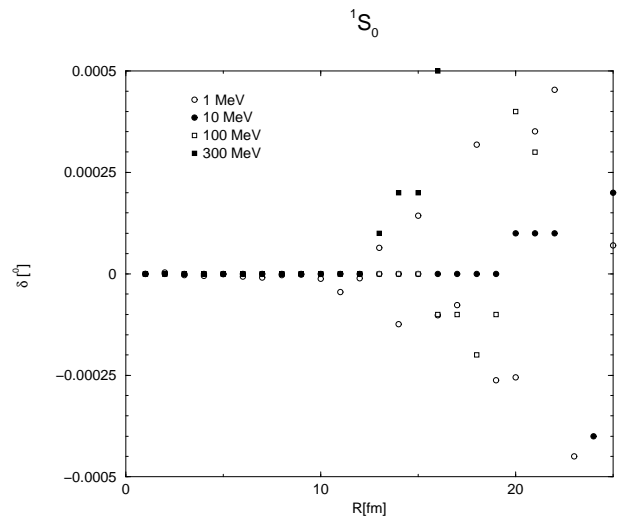, height = 8cm}
\end{turn}
\caption{Phase shifts for the pure Coulomb potential at $E_{\rm lab} =
1,10,100,300\,$MeV as a function of the matching radius $R$.}
\label{pureCoulomb}
\end{center}
\vspace{-1.5cm}
\end{figure}

\begin{figure}[htb]
\begin{center}
%\hspace{-1cm}
\begin{turn}{270}
\epsfig{file=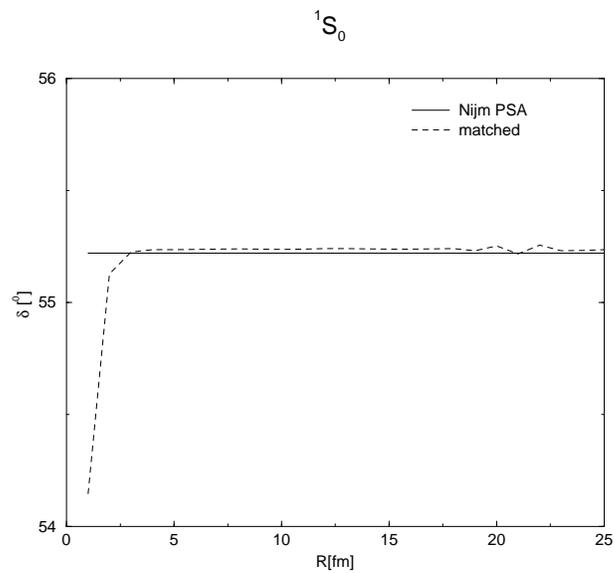, height = 8cm}
\end{turn}
\caption{$^1S_0$ phase shift for Reid93 plus Coulomb potential at
$E_{\rm lab} = 10\,$MeV (solid line) for varying matching radius
(dashed line).}
\label{ReidCoulomb}
\end{center}
\vspace{-1.5cm}    
\end{figure}

\medskip \noindent
The only remaining difficulty is the determination of the
matching radius $R$, because the
given solution is wrong as long as (\ref{cond}) is not valid. On the
other hand, it is not possible to extend $R$ to arbitrarily 
large values, because the 
cosine in eq.(\ref{Coulombpot}) will cause rapid
oscillations. Therefore,  some test calculations are necessary.
First, we put a pure Coulomb potential in our framework. In this
particular case, the phase shift defined in eq.(\ref{ucwave}) should
be zero and $R$ should allowed to be arbitrarily small (but equal to
zero). Nevertheless 
beyond some radius, the abovementioned oscillations are expected. 
In Fig.\ref{pureCoulomb} we have plotted the phase shift for several 
lab energies in the $^{1}S_{0}$ wave.
Second, we took the well known Reid93~\cite{Reid} potential for 
proton-proton scattering  as $V_{\rm strong}$ and re-did the procedure. 
In Fig.\ref{ReidCoulomb} we confer our calculation 
for different $R$ with the published phase shifts of the Reid93-potential
at a fixed energy, $E_{\rm lab} = 10\,$MeV.
As a third check, we compared an r-space solution for a Fourier
transformed one-pion-exchange potential of the
Bonn-type~\cite{Bonn} (dipole cutoff) and the standard Coulomb potential
 with the same one-pion-exchange in our framework.
The checks showed the expected behavior and it seems to be a good
choice to define $R = 10\,$fm in all partial waves  (which is in
accordance  with a Fourier transformation of our strong potential as
well) as it has been done for the Bonn potential~\cite{Hai}. 
For $R$ less than 5-7~fm, the strong potential is still present, for 
$R$ bigger than 15~fm, the oscillations can not be neglected.

\subsection{Range expansion}
\label{sec:range}
It is of particular interest to consider the scattering lengths and
effective range parameters. The effective range expansion takes
the form (written here for a genuine partial wave)
\beq\label{ERE}
p\, \cot(\delta) = -\frac{1}{a} + \frac{1}{2}\, r \, p^2
+ v_2 \, p^4 + v_3 \,p^6 + v_4 \, p^8 + {\cal O}(p^{10})~,
\eeq
with $p$ the nucleon cms momentum, $a$ the scattering length and
$r$ the effective range. It has been stressed in ref.\cite{CH}
that the shape parameters $v_i$ are a good testing ground for the range
of applicability of the underlying EFT since a fit to say the
scattering length and the effective range at NLO leads to predictions
for the higher order range parameters $v_i$. 

\medskip\noindent
We now turn to the proton--proton scattering length. It is well--known that
the strong hadronic part $a_{pp}$ can only be separated from the one in the presence
of the Coulomb interaction $a_{pp}^C$ in a regularization scheme dependent way.
For a cut--off field theory employing a sharp cut-off as done here, one
expects a relation of the form,
\beq\label{appC}
{1 \over a_{pp} (\Lambda)} = {1 \over a_{pp}^C} + \alpha \, 
\left[ F\left(\ln {\Lambda \over m}\right) + {\rm const.} \right]~.
\eeq
Some explicit examples for the function in the square brackets
in a theory where the pions are integrated out are given in ref.\cite{RK} (see
also~\cite{barry}). We refrain from further discussing the explicit
cut--off dependence of this relation but will  come back to this when
we discuss the numerical results. It is also important to stress that within
the EFT, only a certain range of cut--off values is allowed and thus the 
often made statement, that the separation of the hadronic  scattering
length $a_{pp}$ from $a_{pp}^C$ is completely arbitrary, does not hold.

%%%%%%%%%%%%%%%%%%%%%%%%%%%%%%%%%%%%%%%%%%%%%%%%%%%%%%%%%%%%%%%%%%%%%%%%%%%%%%%%
\section{Renormalized potential}
\def\theequation{\arabic{section}.\arabic{equation}}
\setcounter{equation}{0}
\label{sec:pot}

In this section we collect the explicit formulae for the various
pieces of the potential expressed in terms of renormalized quantities.
The infinities appearing in the TPE diagrams are treated exactly in
the same manner as in ref.\cite{EGMII} and we refrain from repeating
these arguments here.

\medskip\noindent
Consider first the one--pion exchange (OPE). Its contribution
is given in the standard Yukawa form. More precisely, for the
$nn$ and the $pp$ systems, only the neutral pion can be exchanged
whereas for the $np$ $T=1$ case, we have a superposition of neutral
and charged pion exchanges,
\beqa
V^{(0)}_{{\rm OPE},pp} =
V^{(0)}_{{\rm OPE},nn}
&=& - \left( {g_A\over 2F_\pi}\right)^2 \, 
\vec{\tau}_1 \cdot \vec{\tau}_2 \,\, {\vec{\sigma}_1 \cdot \vec{q} \,
\vec{\sigma}_2 \cdot \vec{q} \over \vec{q}\,^2 + \Mpnz} ~, \\
V^{(0)}_{{\rm OPE},np} &=& - \left( {g_A\over 2F_\pi}\right)^2 \, 
\vec{\tau}_1 \cdot \vec{\tau}_2 \, \vec{\sigma}_1 \cdot \vec{q} \,
\vec{\sigma}_2 \cdot \vec{q} \, \biggl( {2 \over \vec{q}\,^2 + \Mppz}
- {1 \over \vec{q}\,^2 + \Mpnz} \biggr) ~, 
\eeqa
with $\vec{q} = (\vec{p}\,' -\vec{p}\,)$ the exchanged momentum in the
centre-of-mass system, where $\vec{p}$ and $\vec{p}\,'$ denote
the cms momentum of the incoming and the outgoing two--nucleon pair, in
order. For later comparison, we also give the isospin symmetric one--pion
exchange potential, it reads
\beq
V^{(0)}_{{\rm OPE}, {\rm sym}} = - \left( {g_A\over 2F_\pi}\right)^2 \, 
\vec{\tau}_1 \cdot \vec{\tau}_2 \, {\vec{\sigma}_1 \cdot \vec{q} \,
\vec{\sigma}_2 \cdot \vec{q} \over \vec{q}\,^2 + \Mpz} ~,
\eeq
with
\beq\label{avmass}
\Mp = \frac{2}{3} \Mpp +  \frac{1}{3} \Mpn = 138.03~{\rm MeV}~,
\eeq
the average pion mass. So far, we have expressed the pion--nucleon
coupling in terms of the Lagrangian parameters $g_A$ and $F_\pi$.
With the help of the Goldberger--Treiman relation $g_{\pi NN} = g_A
m_N/F_\pi$ it is then possible to write down these coupling in a more
familiar form. As explained above, we do not need to worry about possible
isospin breaking effects in $g_{\pi NN}$. They are not only suppressed
by the power counting but also  known to be small in the most recent
phase shift analysis~\cite{vKNij}.

\medskip\noindent
Consider now the TPE potential. At NLO, for the case of equal pion
masses, we can use eq.(2.12) of ref.\cite{EGMII} with the average
pion mass as defined in eq.(\ref{avmass}). We give this expression
here for completeness,
\beqa\label{TPEP}
V_{2\pi ,{\rm sym}}
&=& - \frac{ \vec{\tau}_1 \cdot \vec{\tau}_2 }{384 \pi^2 F_\pi^4}\,
L(q) \, \biggl\{4M_\pi^2 (5g_A^4 - 4g_A^2 -1) + q^2(23g_A^4 - 10g_A^2 -1)
+ \frac{48 g_A^4 M_\pi^4}{4 M_\pi^2 + q^2} \biggr\}\no \\
&& - \frac{3 g_A^4}{64 \pi^2 F_\pi^4} \,L(q)  \, \biggl\{
\vec{\sigma}_1 \cdot\vec{q}\,\vec{\sigma}_2\cdot\vec{q} - q^2 \, 
\vec{\sigma}_1 \cdot\vec{\sigma}_2 \biggr\} + P(\vec{k}, \vec{q}\,)~,
\\
L(q) &=& \frac{1}{q}\sqrt{4 M_\pi^2 + q^2}\, 
\ln\frac{\sqrt{4 M_\pi^2 + q^2}+q}{2M_\pi}~,
\eeqa
and we have set $q \equiv |\vec{q}\,|$. Obviously, the first term  of
the NLO isospin symmetric TPEP in eq.(\ref{TPEP}) is isovector while the
second one is isoscalar. The polynom $P(\vec{k}, \vec{q}\,)$ can be
absorbed in the isospin symmetric dimension two four--nucleon contact
interactions up to some small finite shifts as discussed in
the appendix (in what follows, we will ignore this effect). The pion mass difference
in the TPE can be incorporated along the lines outlined in ref.\cite{FvK}.
For completeness, we repeat here the basic steps of that paper and
bring the potential in the form appropriate to our discussion. It is
most convenient to consider the isoscalar and isovector TPE piece
separately,
\beq
V_{2\pi} = V_{2\pi}^0 + V_{2\pi}^1 \, \vec{\tau}_1 \cdot
\vec{\tau}_2~.
\eeq
For the isoscalar part, we can express the TPE as
\beq
 V_{2\pi}^0 = \frac{2}{3} V_{2\pi}^0 (\Mpp, \Mpp ) +
 \frac{1}{3} V_{2\pi}^0 (\Mpn, \Mpn )~,
\eeq
where the arguments refer to the masses of the two exchanged pions. We
remark that due to charge conservation, only similar pions can be
exchanged in this case. This explicit form is used in the numerical
evaluation with $ V_{2\pi}^0$ as given in eq.(\ref{TPEP}). It is
also instructive to perform a Taylor expansion of this expression
around the average pion mass,
\beqa 
\quad &\,&  \frac{2}{3} V_{2\pi}^0 (\Mpp, \Mpp ) + \frac{1}{3} V_{2\pi}^0 (\Mpn,
\Mpn ) \nonumber \\
&=& \frac{2}{3} \biggl( V_{2\pi}^0 (\Mp, \Mp ) + 2 {\partial V^0_{2\pi}\over
\partial \Mpp}\biggl|_{\Mpp=\Mp} (\Mpp - \Mp) + {\cal O}(\alpha^2 )
\biggr)\nonumber \\
&+& \frac{1}{3} \biggl( V_{2\pi}^0 (\Mp, \Mp ) + 2 {\partial V^0_{2\pi}\over
\partial \Mpn}\biggl|_{\Mpn=\Mp} (\Mpn - \Mp) + {\cal O}(\alpha^2 )
\biggr) \nonumber \\
&=&  V_{2\pi}^0 (\Mp, \Mp ) + \frac{4}{3} {\partial V^0_{2\pi}\over
\partial \Mpp}\biggl|_{\Mpp=\Mp} \frac{1}{3} (\Mpp-\Mpn) + \frac{2}{3}
{\partial V^0_{2\pi}\over \partial \Mpn}\biggl|_{\Mpn=\Mp}
\biggl(-\frac{2}{3} \biggr)
(\Mpp - \Mpn) + {\cal O}(\alpha^2 )  \nonumber \\
&=&  V_{2\pi}^0 (\Mp, \Mp )   + {\cal O}(\alpha^2 )~,
\eeqa
i.e. to order ${\cal O} (\alpha)$  there is no isospin breaking in the
isoscalar TPE. For the isovector TPE, we have the general structure
\beqa\label{TPE1}
V_{2\pi}^1 &=& \tau_1^3 \, \tau_2^3 \, V_{2\pi}^1 (\Mpp, \Mpp )
+ (\vec{\tau}_1 \cdot \vec{\tau}_2 -\tau_1^3 \, \tau_2^3 ) \,
V_{2\pi}^1 (\Mpp, \Mpn )  \nonumber \\
&=& \left\{ \begin{array}{ll}
V_{2\pi}^1 (\Mpp , \Mpp ) & {\rm for}~~pp~~{\rm and}~~nn \\
2V_{2\pi}^1 (\Mpp , \Mpn ) - V_{2\pi}^1 (\Mpp , \Mpp ) 
& {\rm for}~~np, T = 1 \end{array}\right. \quad .
\eeqa
Clearly, for the $pp$ (and the $nn$) system, we have to use the
charged pion mass. In case of $np$, we use for the numerical evaluation the
expression in eq.(\ref{TPE1}). If one again considers the pion mass
difference as a small effect and Taylor expands around its average
value, one finds that to order $\alpha$ one has to use the neutral
pion mass in $V_{2\pi}^1$ for the $T=1$ $np$ system.

\medskip\noindent
The $\pi\gamma$ exchange diagrams have already been obtained in
ref.\cite{vKNij} and we use the results obtained in that paper
omitting all computational details here (although it should be
stressed that the treatment of the divergent loop integrals has
to be considered approximative, however, for the accuracy of our
calculation it is sufficient). Due to isospin only charged pion
exchange can contribute to the $\pi\gamma$ potential $V_{\pi\gamma}$
and thus it only affects the $np$ system. The potential has the form
\beqa\label{Vpiga}
V_{\pi\gamma} (\vec{q} \, ) &=& -{g_A^2 \over 4F_\pi^2 \Mppz} \,
(\vec{\tau}_1 \cdot \vec{\tau}_2 -\tau_1^3 \, \tau_2^3 ) \,
\vec{\sigma}_1 \cdot \vec{q} \, \vec{\sigma}_2 \cdot \vec{q} \,\, V_{\pi\gamma}
(\beta)~, \nonumber \\
V_{\pi\gamma}(\beta ) &=& {\alpha \over \pi} \biggl[ -{ (1-\beta^2)^2
  \over 2\beta^4 (1+\beta^2) } \ln(1+\beta^2) + {1\over 2\beta^2} 
- {2\bar{\gamma} \over 1+\beta^2} \biggr]~.
\eeqa 
Here, $\beta = |\vec{q} \,|/\Mpp$ and $\bar{\gamma}$ is a
regularization scheme dependent constant. 
Consistent with the renormalization procedure adopted
in ref.\cite{EGMII} (no finite renormalization), 
we set $\bar{\gamma} = 0$ and the expression 
for the potential simplifies accordingly. The analytical form of
$V_{\pi\gamma}$ is similar to the one of the OPEP but it differs in
strength by the factor $\alpha/\pi \simeq 1/400$. Due to this
inherent smallness, one only expects it to have some influence on the 
S--wave scattering length.

\medskip\noindent
Finally, we consider the contact interactions. In the isospin limit and
at leading order (i.e. with no derivatives), we have two independent contact
interactions and seven at NLO (two derivatives). Particular linear
combinations of these terms can be projected onto the appropriate
partial waves, specifically the two S--waves ($^1S_0, ^3S_1$), the
four P--waves ($^1P_0,^3P_j$) and the mixing parameter $\epsilon_1$ of
the coupled $^3S_1 - ^3D_1$ system (for $T=0$). For the two nucleon
systems with $T = 1$, one has of course only the singlet S--wave
and the three triplet P--waves. In 
case of CIB and CSB, each of the next--to--leading order 
S--wave contact terms consists actually of
three different contributions, namely
the isospin symmetric part $C_{\rm sym}$ (with two derivatives) as defined in
 ref.\cite{EGMII}, the CIB part $\sim \tau_1^3 \, \tau_2^3$ and,
the CSB part $\sim (\tau_1^3 + \tau_2^3)$~. Four the P--waves, we only
have the two--derivative isospin symmetric contact terms since the equivalent
CIB and CSB contact interactions only appear at two orders higher.
Therefore, we concentrate on the S--waves.
The evaluation of these interactions in the different isospin 
channels leads to the following contribution to the various NN potentials
\beqa
C_{pp} &=& C_{\rm sym} + C_{\rm CIB} + C_{\rm CSB}~,\\
C_{nn} &=& C_{\rm sym} + C_{\rm CIB} - C_{\rm CSB}~,\\
C_{np} &=& C_{\rm sym} - C_{\rm CIB}~.
\eeqa  
As long as  only the $pp$ and the $np$ phase shift are used to
pin down the parameters accompanying these contact interactions,
one can extract from the contact interactions
\beqa
C_{pp} - C_{np} &=& 2C_{\rm CIB} + C_{\rm CSB}~, \nonumber\\
C_{pp} + C_{np} &=& 2C_{\rm sym} + C_{\rm CSB}~. 
\eeqa 
Only if one  has access to a third observable, a separation of these
three different contributions is possible. In the $^1S_0$ channel, 
the $nn$ scattering length and the effective range come into play. 
These can be measured by various methods (as discussed below) and
for that particular partial wave, we can then separate the isospin
symmetric, CIB and CSB contributions to the coupling constant $C$ via
\beqa\label{CCSB}
C_{\rm CSB} = \frac{1}{2} (C_{pp} - C_{nn})~, \\
C_{\rm CIB} = \frac{1}{2} (C_{pp} - C_{np} - C_{\rm CSB})~.
\label{CCIB}
\eeqa
This finishes our formal discussion and we now turn to the results.

\section{Results and discussion}
\def\theequation{\arabic{section}.\arabic{equation}}
\setcounter{equation}{0}
\label{sec:res}

\subsection{Fitting procedure}

First, we must fix the parameters. Throughout, we work with 
$g_A = 1.26$, $m_p = 938.27\,$MeV, $m_n = 939.57\,$MeV,
$\Mpn = 134.98\,$MeV and $\Mpp = 139.57\,$MeV. 
Next, we describe the fitting procedure to pin down the LECs. We
proceed along the same lines as detailed in~\cite{EGMII}. We perform
two types of fits. The global ones are obtained by fitting to the
phase shifts of the Nijmegen partial wave analysis for the $np$
and the $pp$ case~\cite{Reid}. More precisely, the isospin--symmetric 
LO and NLO LECs are determined from a best fit  to the S--wave $^1S_0$ and the three
triplet P--waves $^3P_{0,1,2}$ at laboratory energies of 1, 5, 10, 25
and 50 MeV and the N$\LOI$~ LECs entirely from the singlet S--wave (for $np$
and $pp$).\footnote{We are aware that in the absence of magnetic
  interactions and vacuum polarisation the first energy in these fits
  is somewhat problematic.}
  Alternatively, for the S--wave we also use the scattering lengths
to determine the isospin breaking LECs. This fit is marked by a ``$\star$''. 
We work with a sharp momentum--space regulator $\Lambda$ in the LS equation. The
errors for the partial waves are taken from ref.~\cite{Reid}. For a certain
range of $\Lambda$, the $\chi^2$ distribution is shallow as expected
in any cut--off EFT. Here, we are only interested in the cut--off
range which allows to {\em simultaneously} fit the $np$ and the $pp$ systems. 
This is the case for the range $300~{\rm MeV} \leq \Lambda \leq
500$~MeV. The upper limit in this range is given by the fit to the
$pp$ partial waves. We remark that as a check we have reproduced the results of
ref.\cite{EGMII} for the $np$ case (for the parameters and cut--offs
used there). The corresponding values for the
LECs are collected in table~\ref{tab:LECvals} for $\Lambda = 500\,$MeV
and the cut--off dependence of the
LO and NLO S--wave LECs is given in table~\ref{tab:1S0LECs}.
\renewcommand{\arraystretch}{1.2}
\begin{table}[htb] 
\begin{center}

\begin{tabular}{||l||c|c|c|c|c||}
  \hline 
 & $^1S_0$ & $^1S_0~\star$ &$^3P_0$ & $^3P_1$ & $^3P_2$  \\ \hline\hline  
 $C^{\rm LO}_{np}$  [10$^4$ GeV$^{-2}$] & $-$0.133 & $-$0.093 & $--$ & $--$ & $--$ \\
 $C^{\rm NLO}_{np}$ [10$^4$ GeV$^{-4}$] &    1.822 & 2.125 & 1.335 & $-$0.394
                                        &    0.191  \\ \hline
 $C^{\rm LO}_{pp}$  [10$^4$ GeV$^{-2}$] & $-$0.129 & $-$0.086 & $--$ & $--$ & $--$ \\
 $C^{\rm NLO}_{pp}$ [10$^4$ GeV$^{-4}$] &    1.822 & 2.125 & 1.335 & $-$0.394
                                        &    0.191  \\ \hline
  \end{tabular}
\caption{Values of the LECs for in the various  partial waves 
 for a  sharp cut--off $\Lambda = 500\,$MeV. \label{tab:LECvals}} 
\end{center}
\end{table}
\begin{table}[htb] 
\begin{center}

\begin{tabular}{||l||c|c|c||}
  \hline 
  $\Lambda$~[MeV] & 300 & 400 & 500 
 \\ \hline\hline  
 $C^{\rm LO}_{np}$  [10$^4$ GeV$^{-2}$] & $-$0.1550 & $-$0.1545 & $-$0.1331 \\
 $C^{\rm NLO}_{np}$ [10$^4$ GeV$^{-4}$] &    1.613  &    1.621  &
 1.822  \\ \hline
 $C^{\rm LO}_{pp}$  [10$^4$ GeV$^{-2}$]  & $-$0.1537 & $-$0.1532 & $-$0.1290 \\
 $C^{\rm NLO}_{pp}$ [10$^4$ GeV$^{-4}$]  &    1.613  &    1.621  &
 1.822  \\
\hline

  \end{tabular}
\caption{Values of the LECs for in the $^1S_0$ partial waves as
 a function of the sharp cut--off. \label{tab:1S0LECs}} 
\end{center}
\end{table}

\subsection{Phase shifts}

\begin{figure}[htb]
\begin{center}
%\hspace{-1cm}
\begin{turn}{270}
\epsfig{file=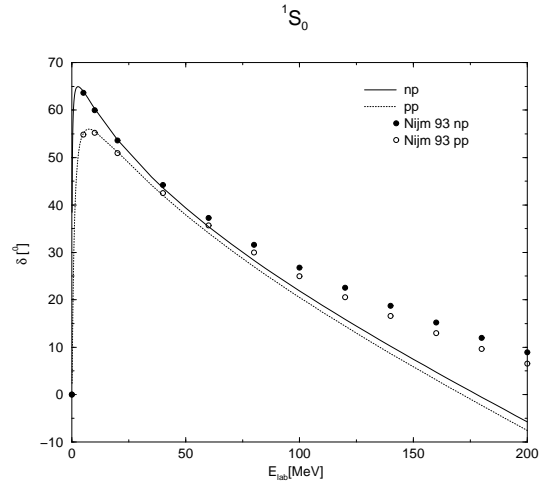, height = 7cm}
\end{turn}
\caption{$^1S_0$ phase shifts for the $np$ and $pp$ systems 
 (solid and dashed line, respectively) im comparison to
  the Nijmegen PSA, filled circles: $np$, open circles: $pp$.}

\label{fig:1S0200}
\end{center}
\end{figure}

\begin{figure}[htb]
\begin{center}
\begin{turn}{270}
\epsfig{file=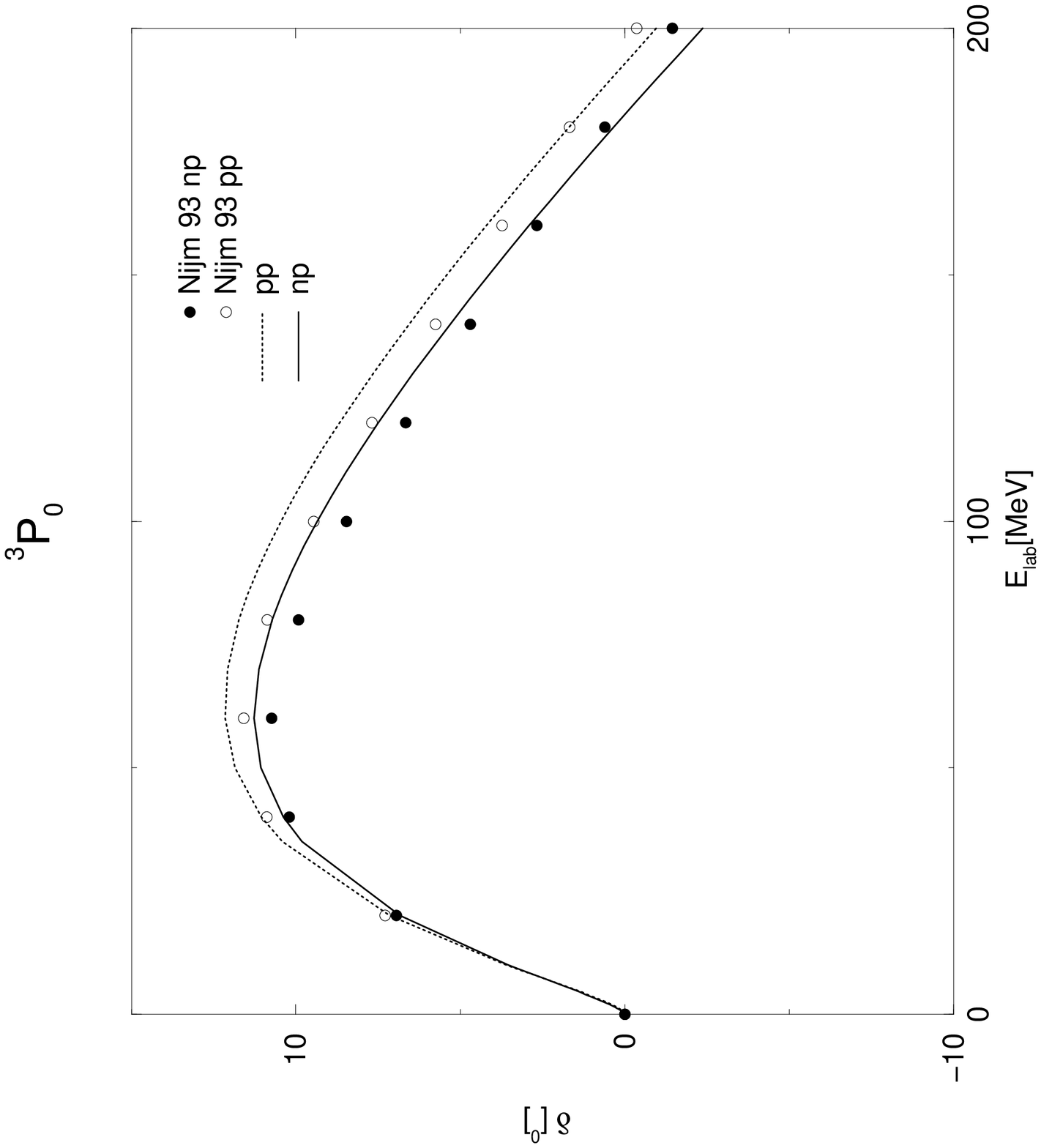, height = 6cm}
\end{turn}
\hspace{1.5cm}
\begin{turn}{270}
\epsfig{file=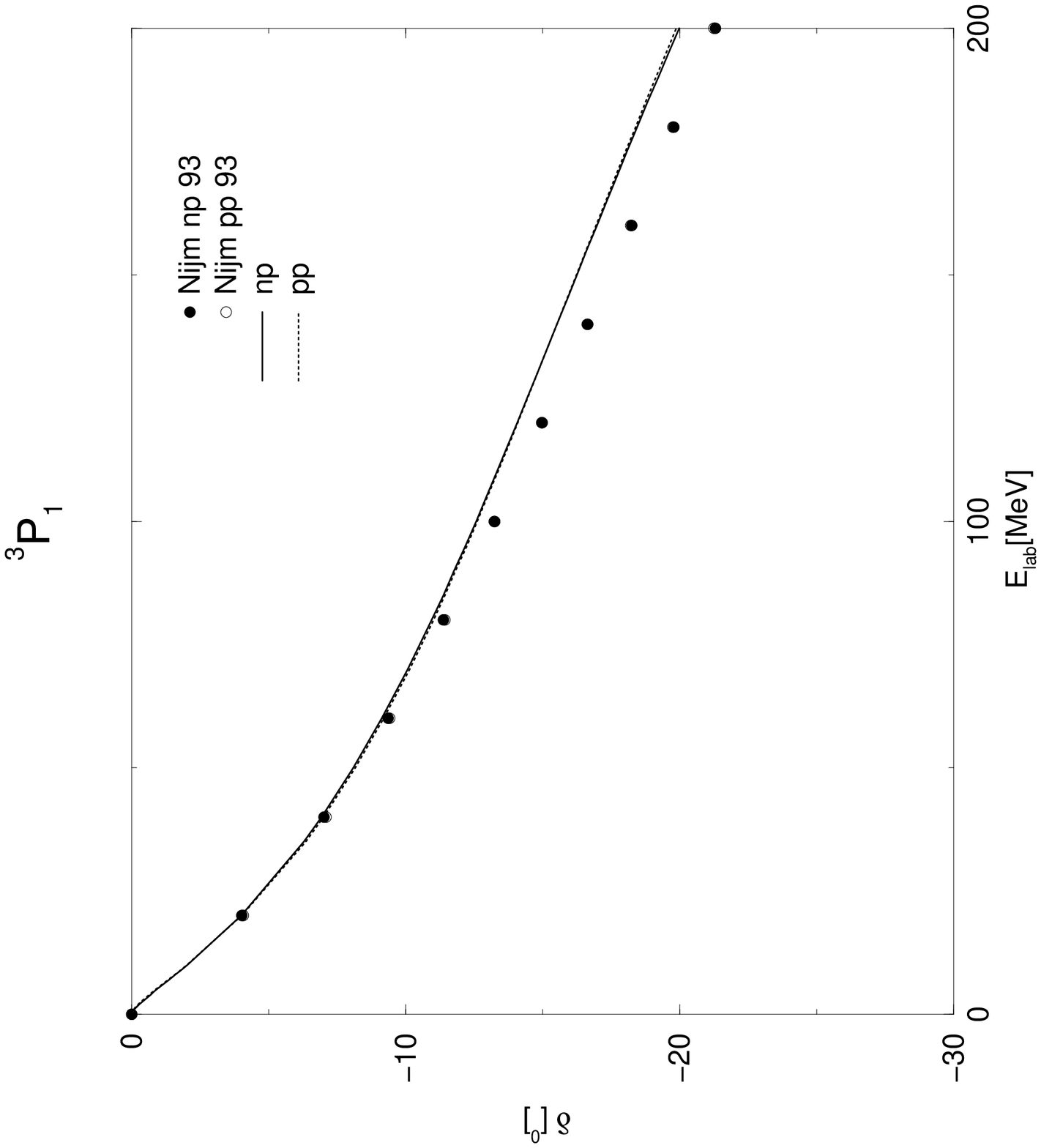, height = 6cm}
\end{turn}
\vspace{1cm}
\begin{turn}{270}
\epsfig{file=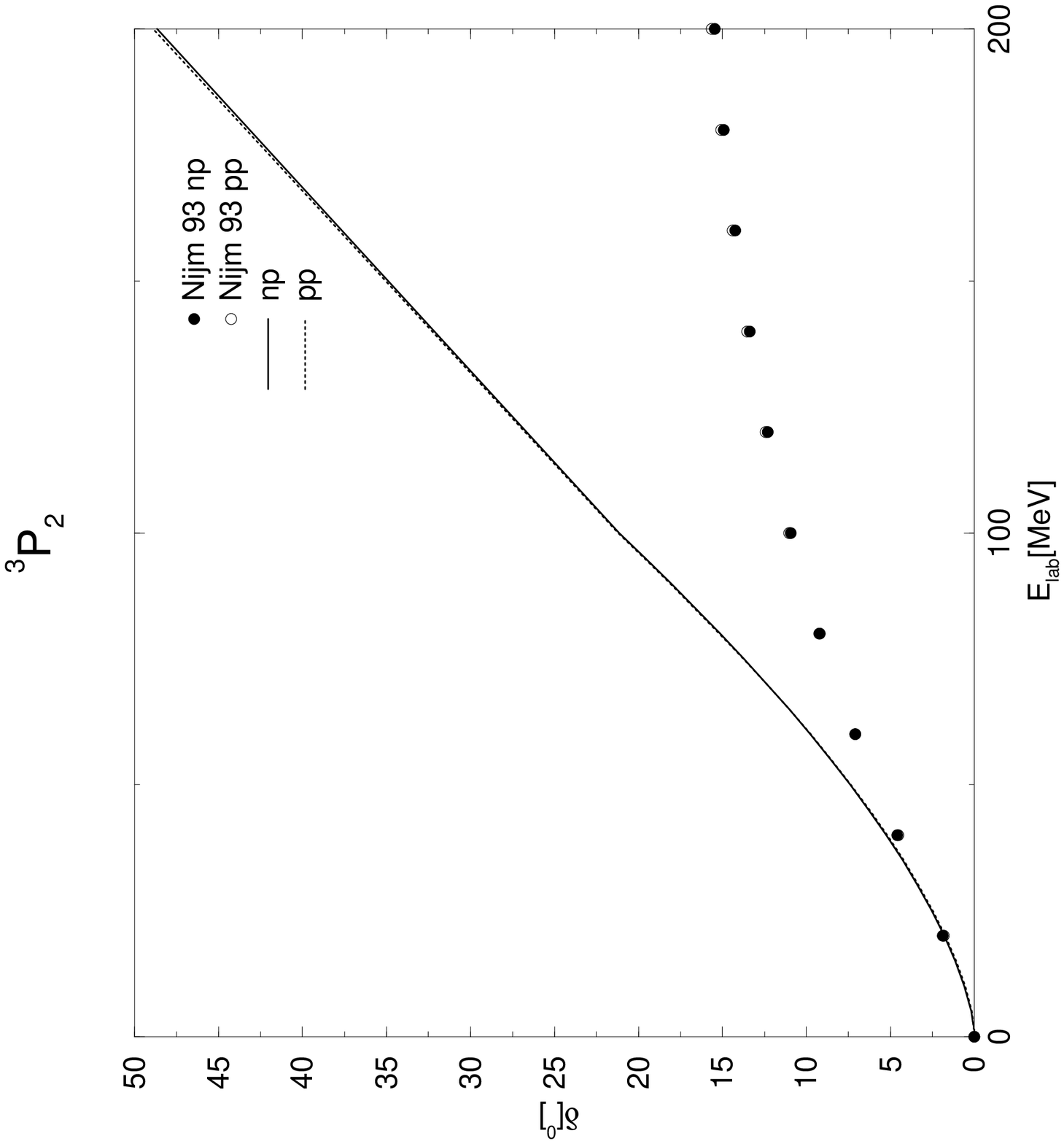, height = 6cm}
\end{turn}
\caption{P--wave phase shifts for the $np$ and $pp$ systems im comparison to
  the Nijmegen PSA. For notations, see Fig.\ref{fig:1S0200}.}
\label{fig:P200}
\end{center}
%\vspace{-0.5cm}    
\end{figure}

\begin{figure}[ht]
\hskip 8.5truecm
\epsfxsize=6truecm
%\epsfysize=1.8in
%\epsffile{dis1.ps}
%\epsffile{fft1.ps}
\begin{turn}{270}
%\epsffile{3F2nloplot.eps}
\epsffile{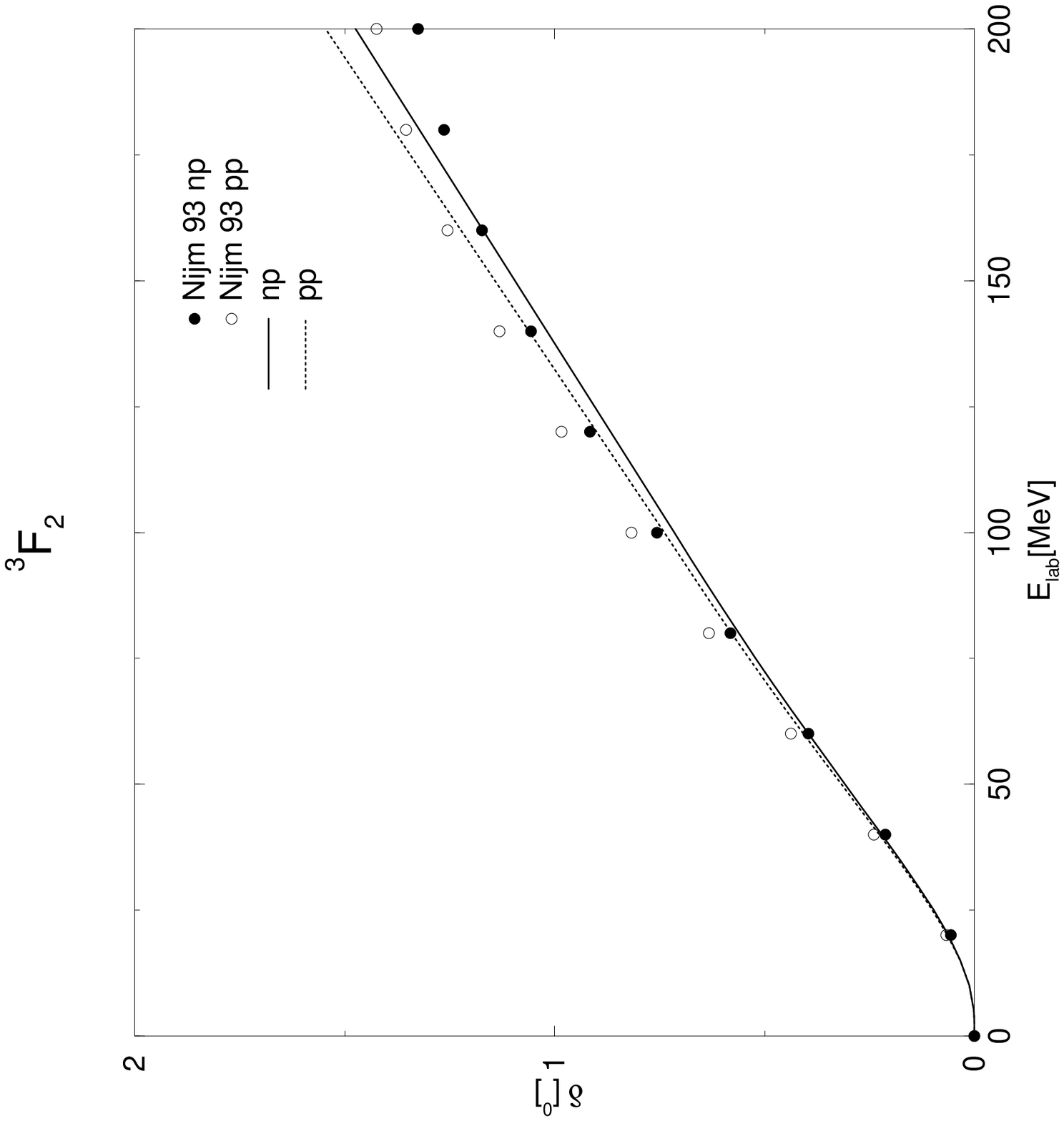}
\end{turn}

%\protect{\hskip -1truein}

\vskip -6truecm
\hskip 1truecm
\epsfxsize=6truecm
\hskip .1truein
\begin{turn}{270}
%\epsffile{1D2nloplot.eps}
\epsffile{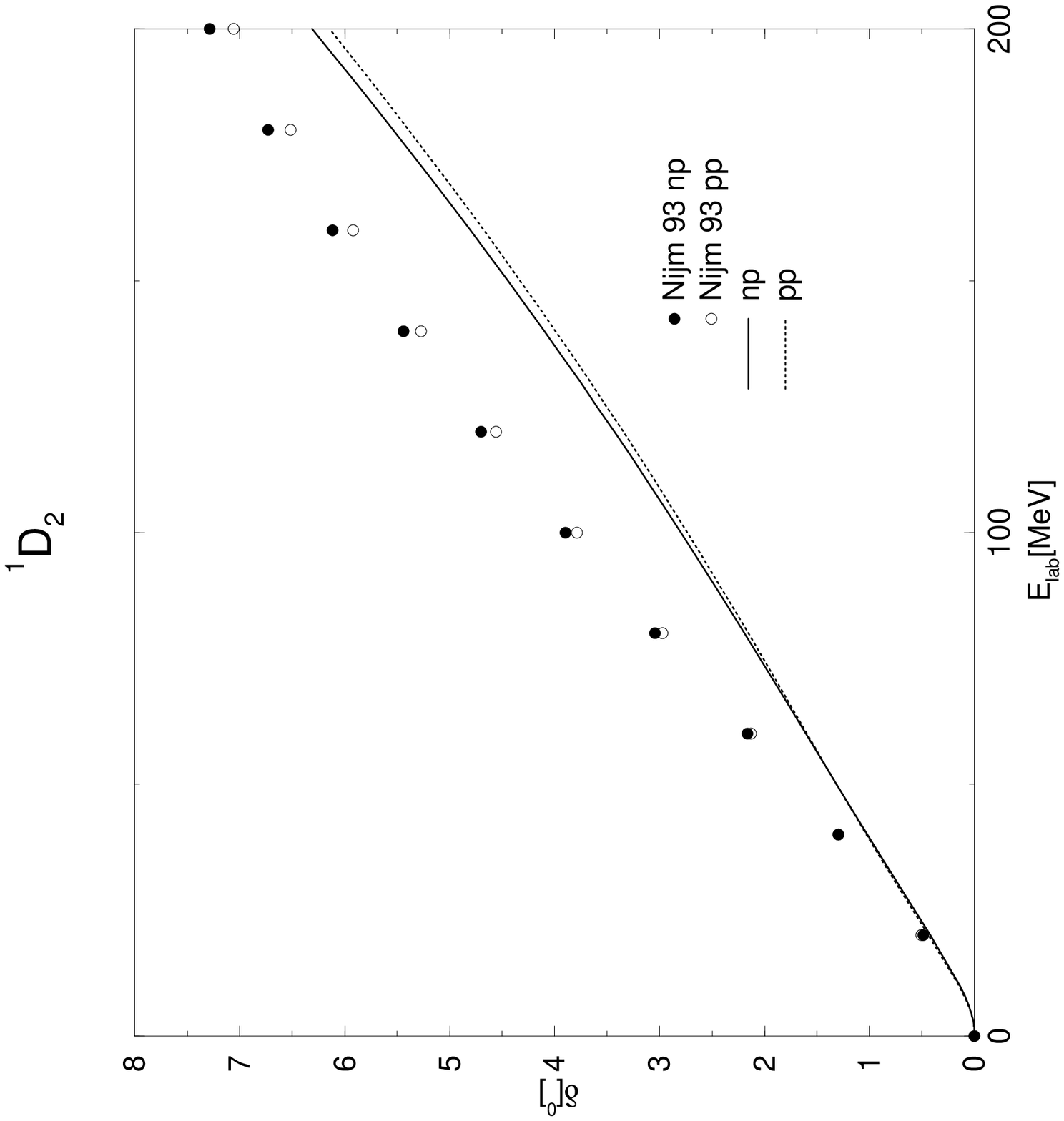}
\end{turn}

\vskip 1truecm

\hskip 8.5truecm
\epsfxsize=6truecm
%\epsfysize=1.8in
%\epsffile{dis1.ps}
%\epsffile{fft1.ps}
\begin{turn}{270}
%\epsffile{1G4nloplot.eps}
\epsffile{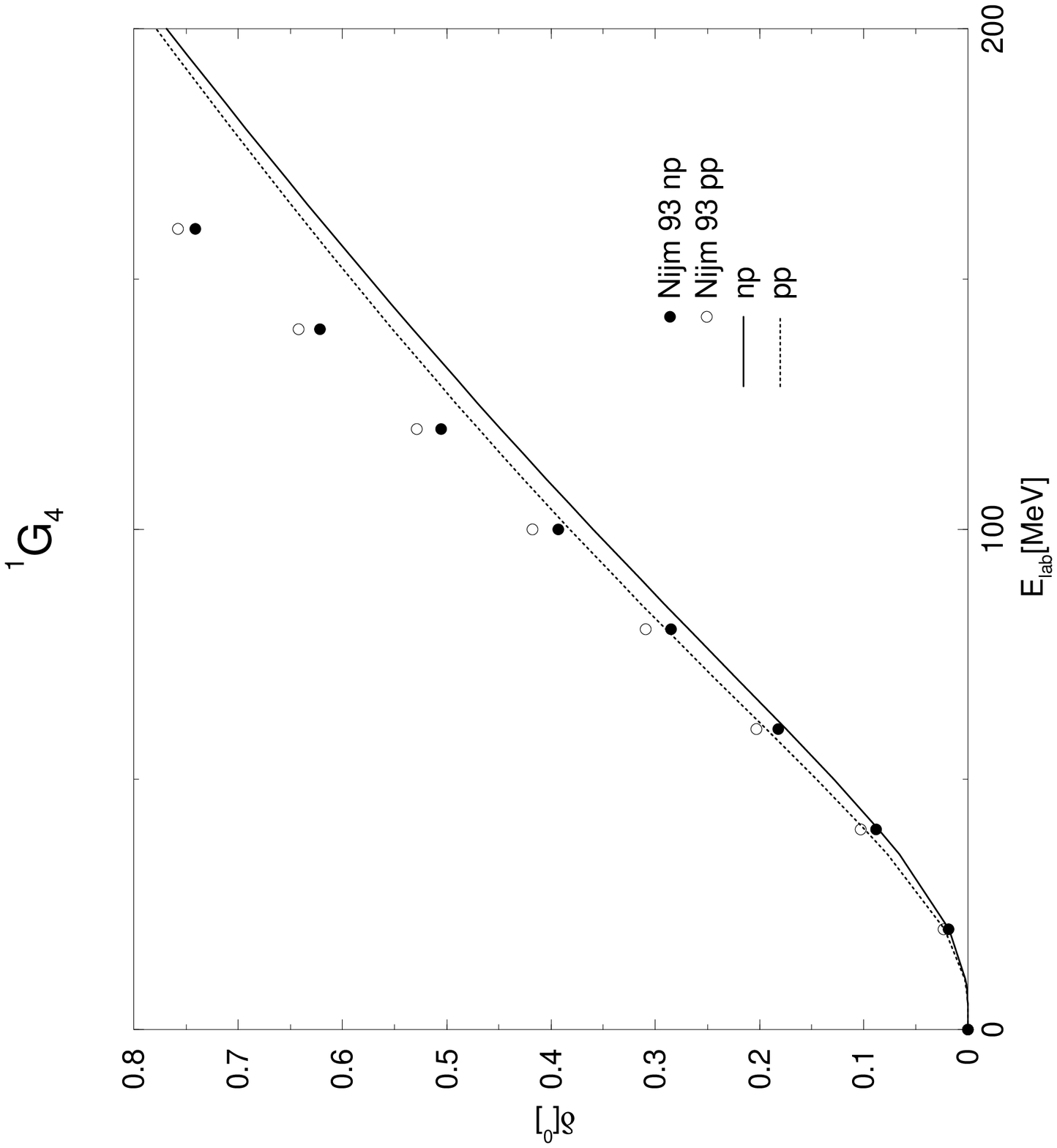}
\end{turn}

%\protect{\hskip -1truein}

\vskip -6truecm
\hskip 1truecm
\epsfxsize=6truecm
\hskip .1truein
\begin{turn}{270}
%\epsffile{3F3nloplot.eps}
\epsffile{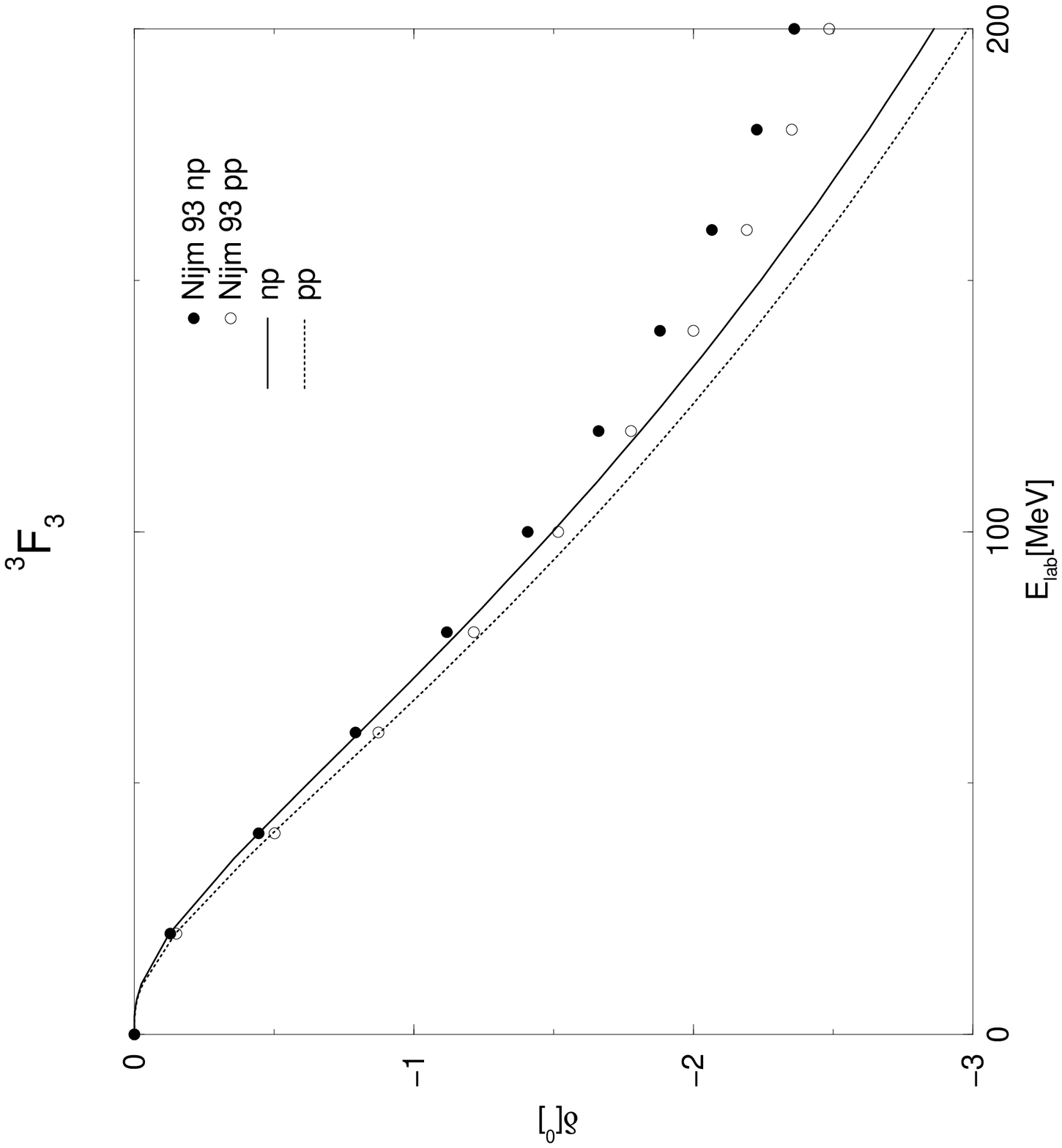}
\end{turn}

\vspace{2cm}
\begin{center}
 \caption{Higher partial waves for the $np$ and $pp$ systems
          in comparison to the Nijmegen PSA. For notations, 
          see Fig.\ref{fig:1S0200}.}           
\label{fig:DFG200} 
\end{center}
\end{figure}

Having determined the LECs, we can now predict the S-- and P--waves
for energies larger than 50~MeV and all other partial waves are
parameter free predictions. To be specific, we employ the cut--off
$\Lambda = 500\,$ MeV. In Fig.\ref{fig:1S0200}, we show the $^1S_0$ partial
wave for energies up to 200 MeV. As noted in ref.~\cite{EGMII}, for 
an improved description of this partial wave up to $E_{\rm lab} = 300$~MeV, one has
to go to NNLO. The triplet P--waves are collected in Fig.\ref{fig:P200}. As already
pointed out in~\cite{EGMII}, the pion mass difference plays an important
role in $^3P_0$ for which we get a good description up to $E_{\rm lab} =
200\,$MeV. The good description of the $^3P_1$ phase up to the
inelastic threshold can be traced back to the dominance of the OPE.
The insufficient description of the $^3P_2$ wave was already be observed
in \cite{EGMII} and can be cured to some extend at N$^2$LO.
Some higher partial waves $(^1D_2,^3F_2,^3F_3,^1G_4)$, which are free 
of tunable parameters at NLO, are shown in Fig.\ref{fig:DFG200}. Depending on the
partial wave, we find satisfactory description of these waves up to
150~MeV. This holds in particular for the mixing parameter
$\epsilon_2$ not shown in the figure. These results can be improved by
going to NNLO, but for our purpose the
accuracy achieved for lab energies up to 150$\ldots$200 MeV is sufficient. In
particular, it allows to study CIB and CSB which is most pronounced in the
threshold region.

\subsection{Charge independence breaking}

\begin{figure}[htb]
\begin{center}
%\hspace{-1cm}
\begin{turn}{270}
\epsfig{file=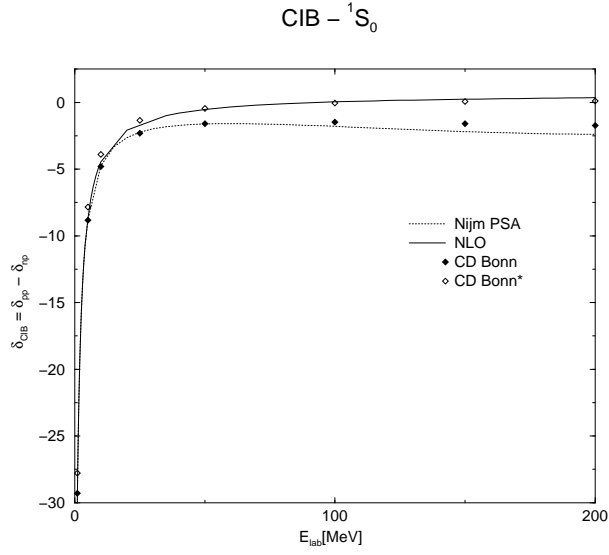, height = 8cm}
\end{turn}
\caption{CIB $^1S_0$ phase shift (solid line) im comparison to
  the Nijmegen PSA (dashed line) and the CD-Bonn potential
  (filled and open squares, as explained in the text).}
\label{fig:1S0CIB}
\end{center}
%\vspace{-0.5cm}    
\end{figure}

\begin{figure}[htb]
\begin{center}
\begin{turn}{270}
\epsfig{file=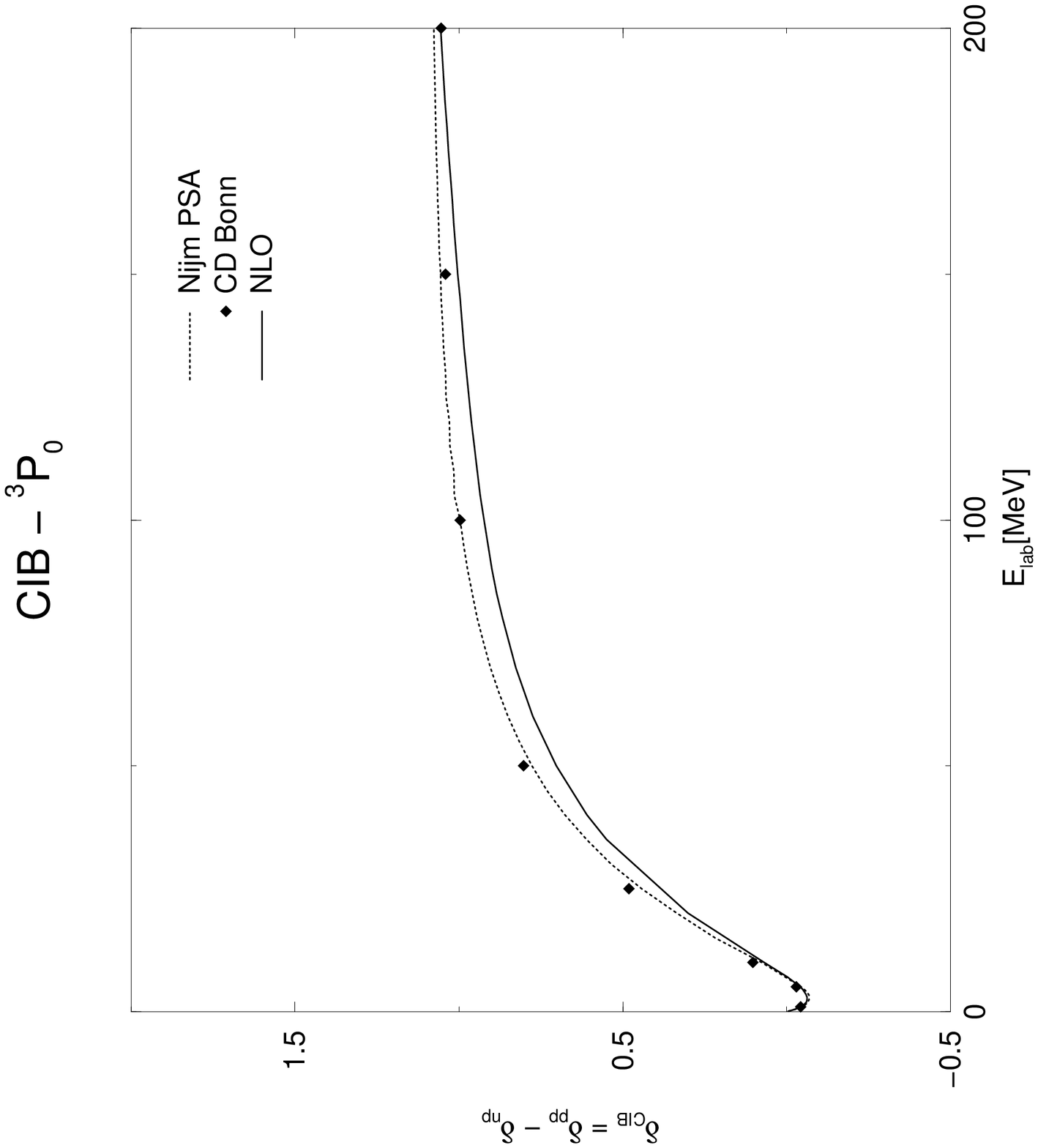, height = 6cm}
\end{turn}
\hspace{1.5cm}

\begin{turn}{270}
\epsfig{file=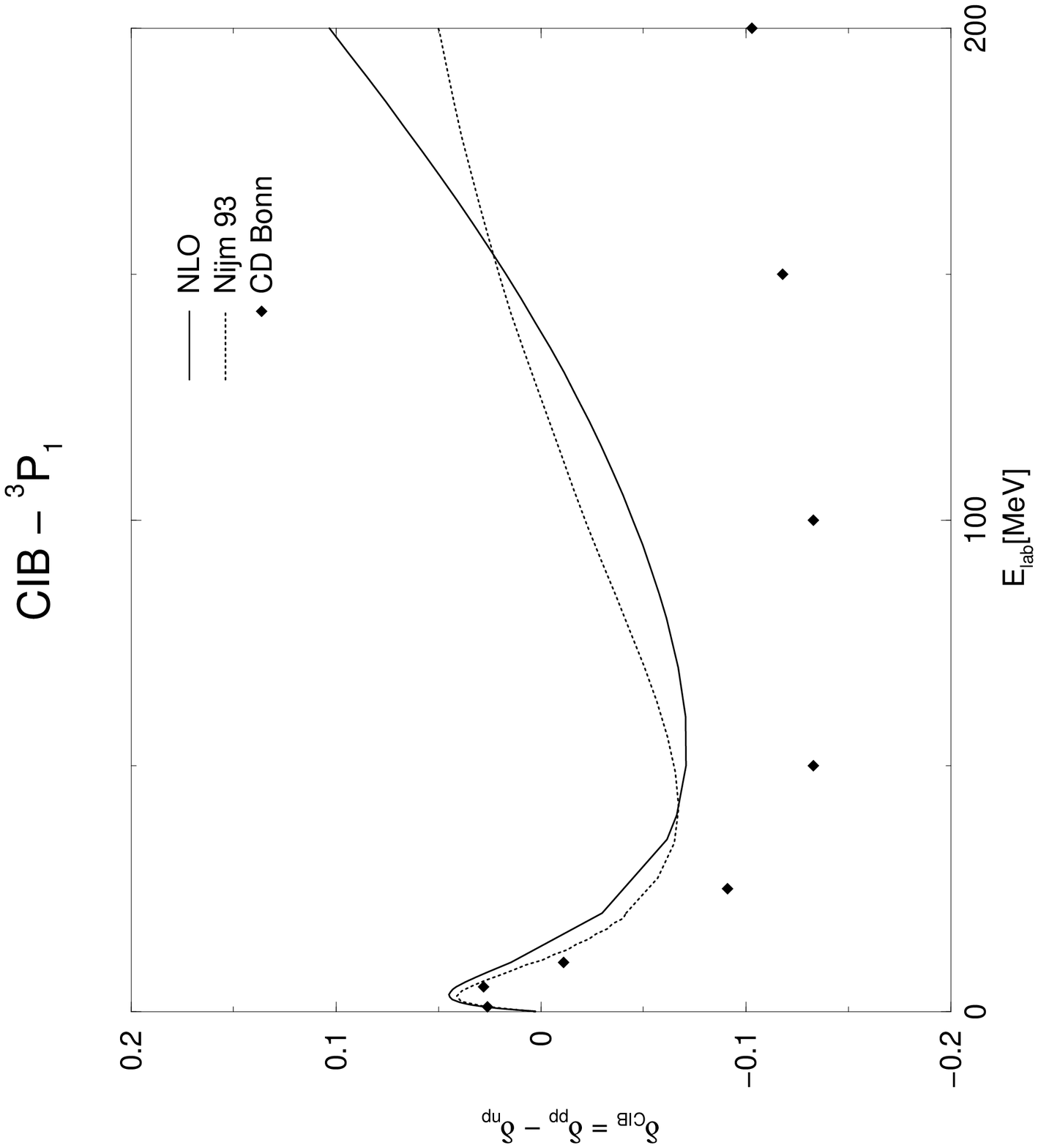, height = 6cm}
\end{turn}
\vspace{1cm}
\begin{turn}{270}
\epsfig{file=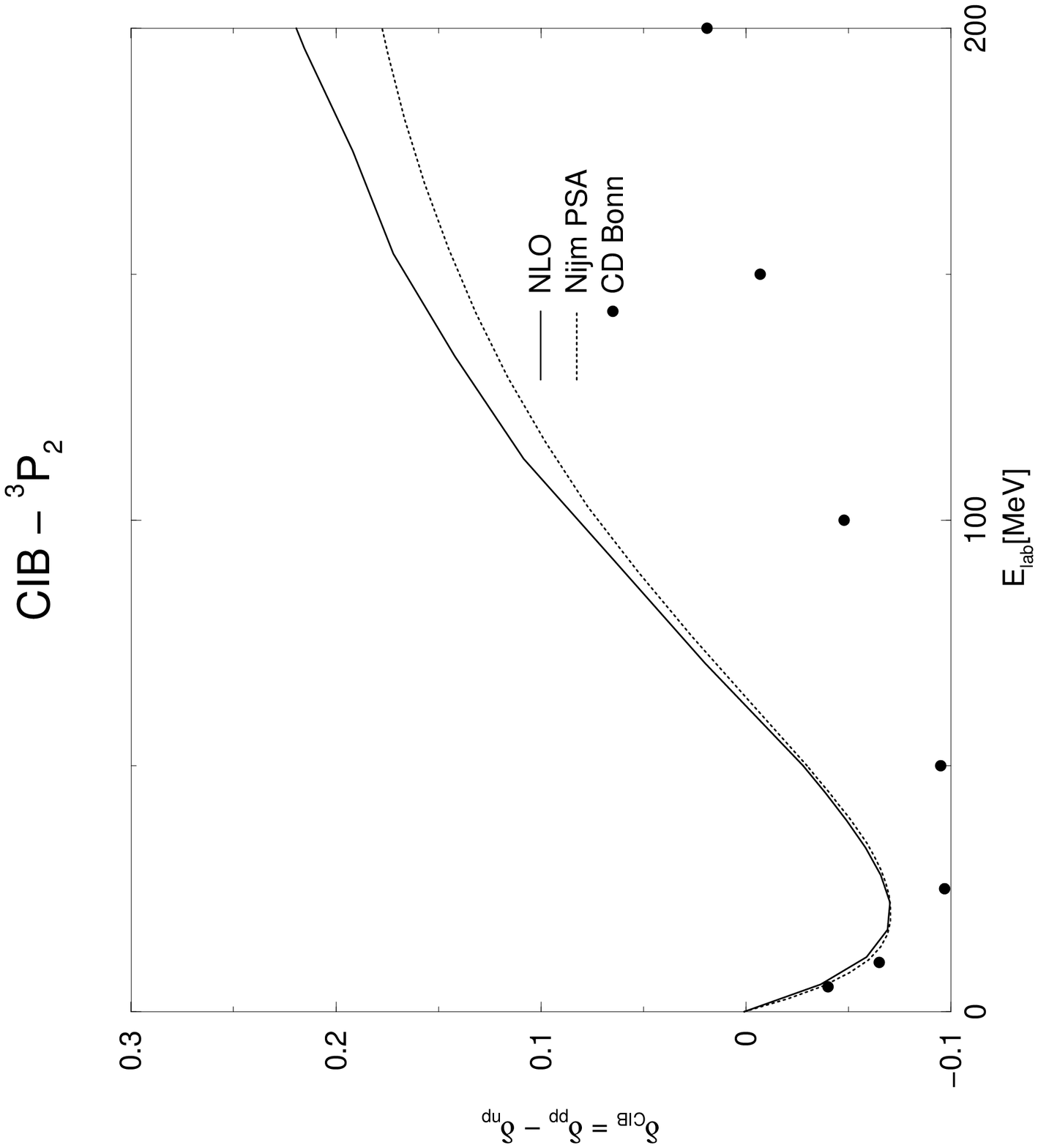, height = 6cm}
\end{turn}
\caption{CIB P--wave phase shifts in comparison to
  the Nijmegen PSA and the CD-Bonn potential.
  For notations, see Fig.\ref{fig:1S0CIB}.}
\label{fig:PCIB}
\end{center}
%\vspace{-0.5cm}    
\end{figure}

\begin{figure}[ht]
\hskip 8.5truecm
\epsfxsize=6truecm
%\epsfysize=1.8in
%\epsffile{dis1.ps}
%\epsffile{fft1.ps}
\begin{turn}{270}
%\epsffile{3F2cibplot.eps}
\epsffile{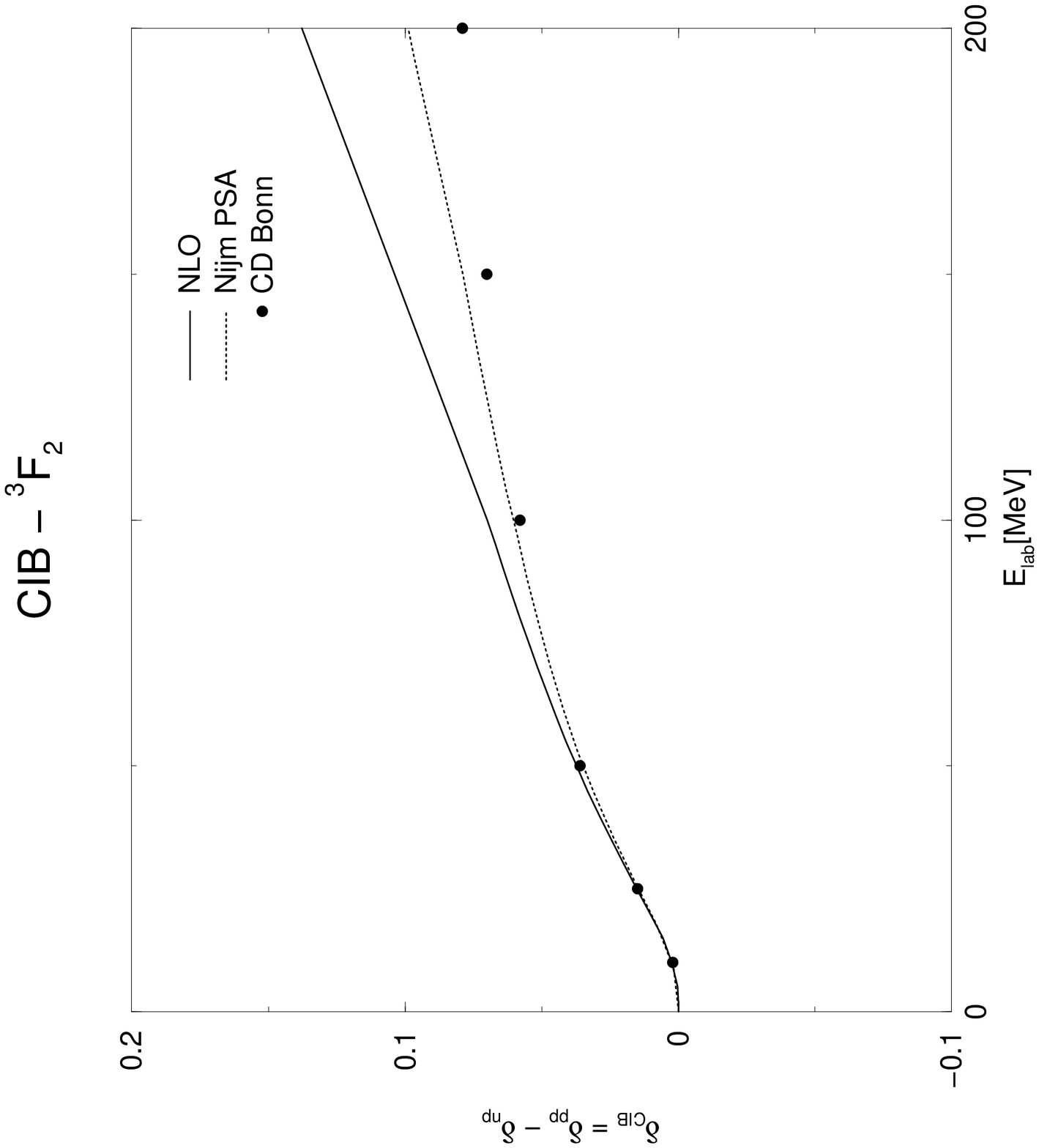}
\end{turn}
%\protect{\hskip -1truein}

\vskip -6truecm
\hskip 1truecm
\epsfxsize=6truecm
\hskip .1truein
\begin{turn}{270}
%\epsffile{1D2cibplot.eps}
\epsffile{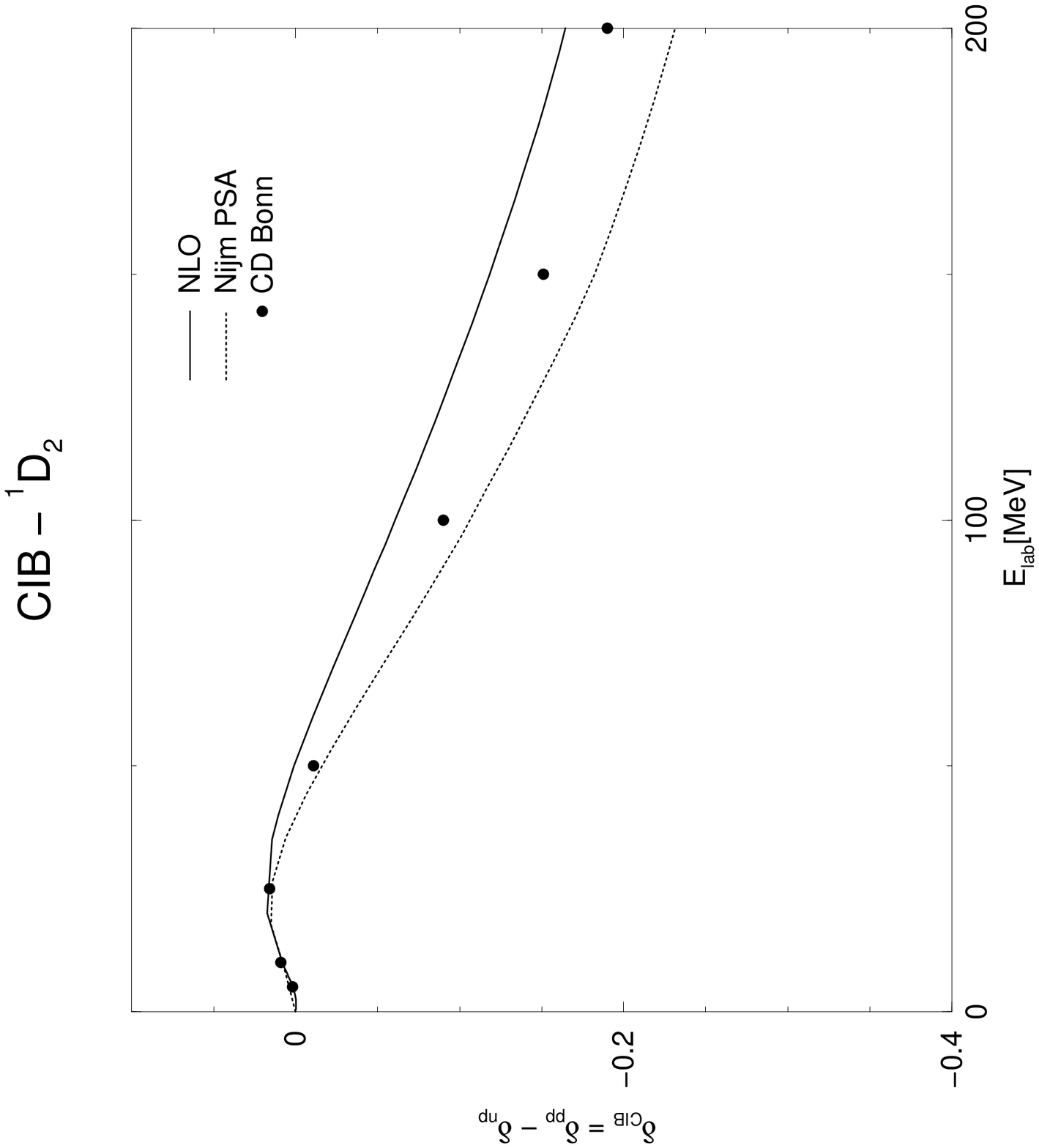}
\end{turn}

\vskip 1truecm
\hskip 8.5truecm
\epsfxsize=6truecm
%\epsfysize=1.8in
%\epsffile{dis1.ps}
%\epsffile{fft1.ps}
\begin{turn}{270}
%\epsffile{1G4cibplot.eps}
\epsffile{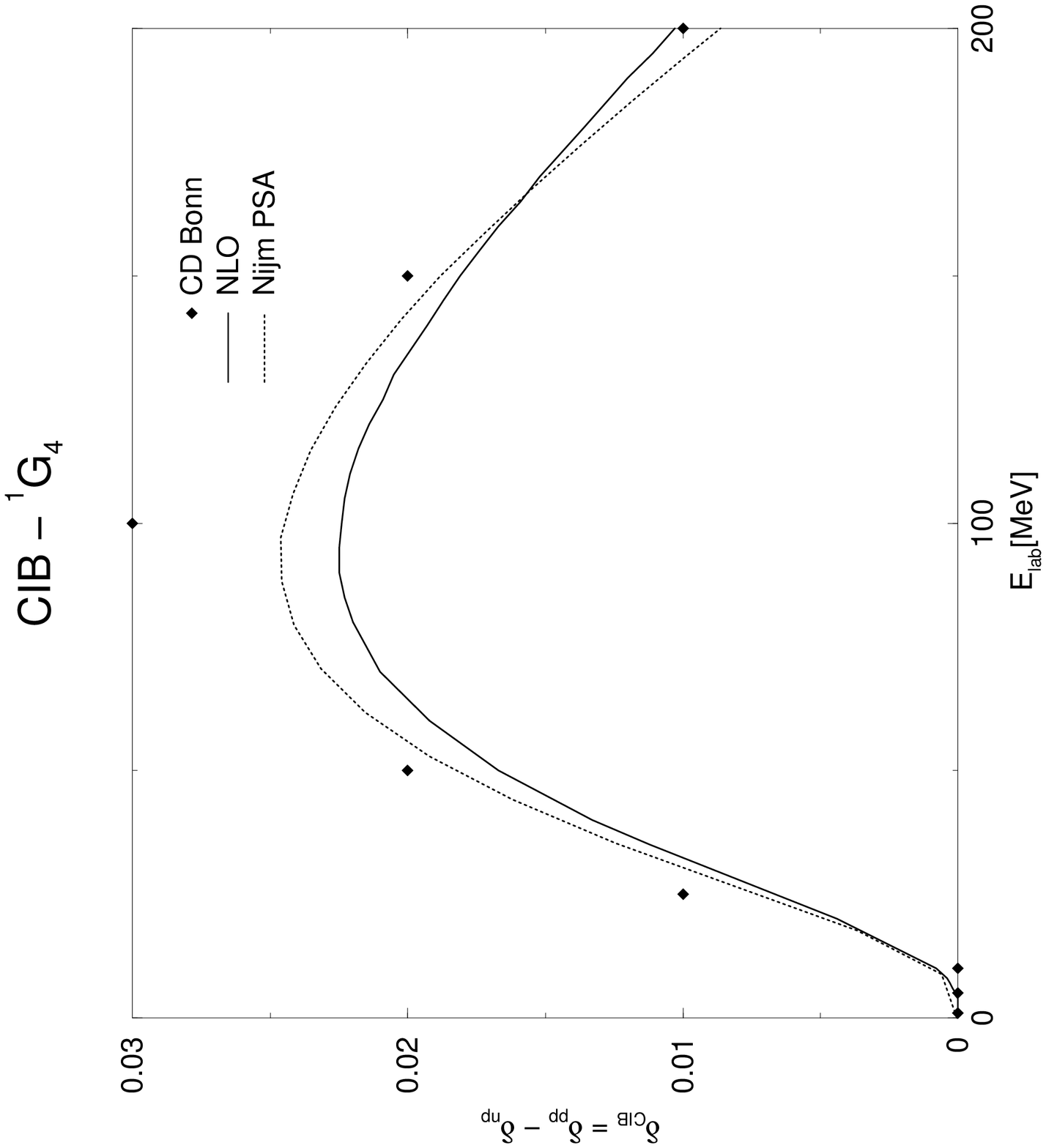}
\end{turn}

%\protect{\hskip -1truein}

\vskip -6truecm
\hskip 1truecm
\epsfxsize=6truecm
\hskip .1truein
\begin{turn}{270}
%\epsffile{3F3cibplot.eps}
\epsffile{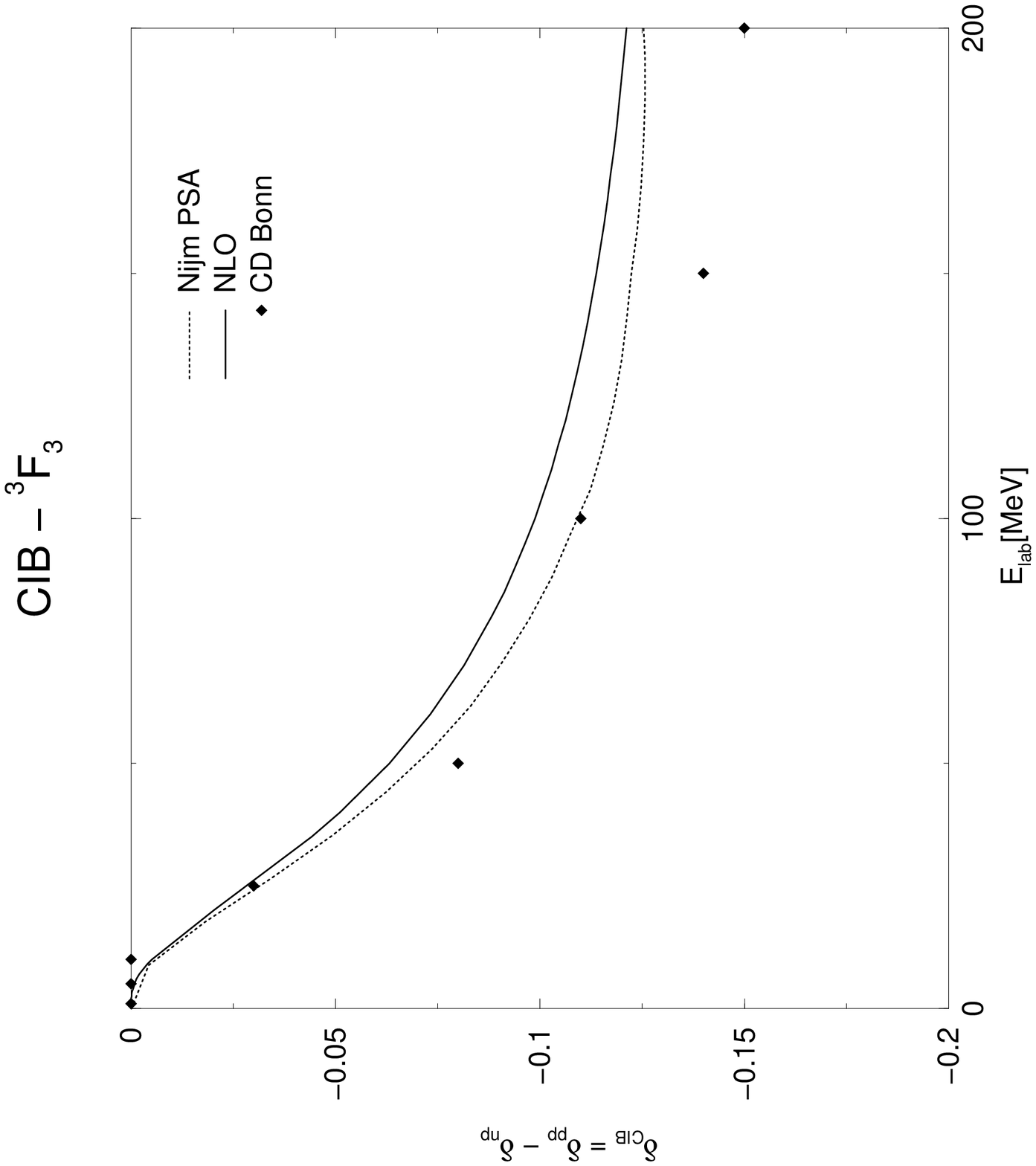}
\end{turn}
\vspace{2cm}
\begin{center}
 \caption{CIB phase shifts in the higher partial waves 
          in comparison to the Nijmegen PSA and the CB-Bonn potential.           
  For notations, see Fig.\ref{fig:1S0CIB}.}
\label{fig:DFGCIB} 
\end{center}
\end{figure}

In the absence of $nn$ data, we define
\beq
X_{\rm CIB} = X_{pp} - X_{np}~, \quad X=\{\delta,a,r,v_1,v_2,v_3\}~,
\eeq
for the pertinent phase shift,  scattering length and range parameters (in a given
partial wave). We show first the energy dependence of $\delta_{\rm CIB}$ in the
various partial waves, in comparison to the Nijmegen PSA and the most recent
results obtained from the CD-Bonn potential. The S-- and P--waves are
collected in Figs.\ref{fig:1S0CIB} and \ref{fig:PCIB}, respectively. 
In the S--wave, our prediction for energies above 50~MeV
shows an interesting difference to the result of the CD-Bonn potential
(filled squares in fig.\ref{fig:1S0CIB}). In that
approach a phenomenological term had to be added to describe correctly the
CIB in the scattering lengths, this affects in particular the CIB difference
above $E_{\rm lab} = 50\,$MeV as depicted by the open squares. The prediction
by ignoring this additional term, whose explicit structure is not
spelled out in ref.\cite{CD-BonnII}, agrees with our full N$\LOI$ result. 
For $^3P_{1,2}$,
our predictions based on the Nijmegen PSA are distinctively different from the
CD-Bonn results, however, these phase differences are in both cases of very
small magnitude. The CIB in the higher partial waves is shown in
Fig.\ref{fig:DFGCIB}, only in $^3F_3$ one finds some difference between our
prediction and the one obtained using the CD-Bonn potential.

\medskip
\noindent
We now turn to the scattering lengths and range parameters.
To be specific, we consider the $^1S_0$ partial wave.
The $np$ and $pp$ scattering lengths as a function of the cut--off
$\Lambda$ are given in table~\ref{tab:arange}. While the result for
$a_{np}$ is, of course, independent of the cut--off (within an
accuracy of about 0.3\%), the $pp$
scattering length varies due to the subtraction ambiguity discussed
in section~\ref{sec:range}. It is, however, gratifying to see that within the range
of allowed cut--offs, this variation is modest, i.e. $|\delta a_{pp} /
a_{pp}| \simeq 0.6~{\rm fm}/17~{\rm fm} \simeq 3\%.$ Consequently, we
get
\beq
a_{\rm CIB} =  6.14 \ldots 6.75~{\rm fm}~,
\eeq
for $\Lambda = 300 \ldots 500\,$MeV.
\renewcommand{\arraystretch}{1.2}
\begin{table}[htb] 
\begin{center}
\begin{tabular}{||l||c|c|c||}
  \hline 
 $\Lambda$~[MeV] & 300 & 400 & 500
 \\ \hline\hline  
 $a_{np}$ [fm]  & $-$23.65 & $-$23.67 & $-$23.71 \\
 $a_{pp}$ [fm]  & $-$17.51 & $-$17.53 & $-$16.96 \\
\hline
  \end{tabular}
\caption{$np$ and $pp$ scattering lengths in  $^1S_0$ partial wave for the range of
cut--off parameters discussed in the text. \label{tab:arange}} 
\end{center}
\end{table}

\noindent
To further dissect the physics underlying CIB, we list in 
table~\ref{tab:CIB} the various contributions to the CIB
in the scattering length and effective ranges due to the
neutron--proton mass difference ($\Delta m_N$), the pion--mass
difference ($\Delta M_\pi$) in OPE and TPE, from $\pi\gamma$ graphs
and due to the contact interactions.
These contributions have been obtained in the following manner: The
total potential is a sum of different terms and to assess the
importance of one particular mechanism, we simply switch that term
off and obtain its contribution as the difference between the full
result and the one in the absence of that term. Due to the inherent
non--linearity of the problem (iteration of the potential,
pseudo--boundstate close to threshold) the sum of the individual
contributions does not give the full result. In the literature one
can find a set of prescriptions to rescale these contributions based 
on an expansion in the inverse of the scattering length. We do not
follow these schemes but stick to the prescription just outlined.
As expected, the pion mass difference is a very important effect, but is
alone insufficient to account for the CIB in the scattering lengths. The TPE
contribution deserves some discussion. It has already been worked out
in ref.\cite{FvK}, where a similar result with an opposite sign was
found. There are several reasons for that disagreement. First of all, the
authors of ref.\cite{FvK} performed all calculations in configuration space
whereas we are working in momentum space. Several zero range terms arising 
by Fourier transformation of the non--polynomial part of eq.(\ref{TPEP}) 
have not been taken into account in ref.\cite{FvK}. Secondly, we are using a
sharp momentum cut--off whereas some smooth local coordinate space
cut--off has been applied in ref.\cite{FvK}. Finally, the charge independent
part of the potential was chosen in \cite{FvK} to be the one of the Argonne 
AV18 potential. In this sense, our description is more consistent and is
entirely based on the potential derived within chiral EFT. Therefore, a
direct comparison with the numerical results of \cite{FvK} should not be made.
We find the same ordering in the importance
of the three types of diagrams, box larger than triangle larger than
football. However, there is also a strong cut--off dependence,
e.g. the box diagram contribution to $a_{\rm CIB}$ changes form
-1.6~fm to -1.1~fm as $\Lambda$ varies from 500 to 300~MeV. 
Such a strong cut--off dependence has, in fact, to be expected, since
neither the contribution to CIB of the complete CIP TPE nor of its 
non--polynomial part, shown in table~\ref{tab:CIB}, correspond to any
observable quantity. Only the complete CIB contribution to phase shifts can be
measured experimentally. The $\pi \gamma$ contribution agrees with the findings in
ref.\cite{vKNij}. Finally, we would like to stress the rather sizeable
contribution of the contact interactions to $a_{\rm CIB}$, which
deserve some study within models of CIB.
\renewcommand{\arraystretch}{1.2}
\begin{table}[htb] 
\begin{center}

\begin{tabular}{||l||r|r|r|r|r||}
  \hline 
 & $\Delta m_N$ & $\Delta M_\pi$ OPE & $\Delta M_\pi$ TPE & $\pi\gamma$
 & contact \\ \hline\hline  
 $a$ [fm]       &    0.1284 &    2.9173 & $-$1.1376 & $-$0.4203 & 4.5083 \\
 $r$ [fm]       & $-$0.0019 & $-$0.0804 &    0.0135 &    0.0066 & $-$0.0739 \\
 $v_1$ [fm$^3$] & $-$0.0018 & $-$0.2684 &    0.0083 &    0.0143 & $-$0.0589 \\
 $v_2$ [fm$^5$] & $-$0.0120 & $-$2.2246 &    0.0541 &    0.0893 & $-$0.4249 \\
 $v_3$ [fm$^7$] & $-$0.0719 & $-$17.800 &    0.2830 &    0.5606 & $-$2.5355 \\
\hline
  \end{tabular}
\caption{Various contributions to CIB in the $^1S_0$ partial wave 
 for a sharp cut--off $\Lambda= 500$~MeV as
 explained in the text. \label{tab:CIB}} 
\end{center}
\end{table}

\subsection{The $nn$ system and charge symmetry breaking}
\label{sec:nn}
\begin{figure}[htb]
\begin{center}
%\hspace{-1cm}
\begin{turn}{270}
\epsfig{file=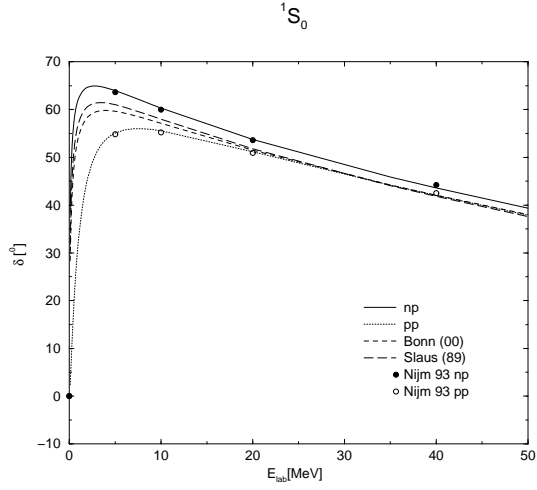, height = 7cm}
\end{turn}
\caption{Range fit for the $nn$ $^1S_0$ phase shift
 based on different scattering lengths $(a_{nn} = -18.9\,$fm: long-dashed
 line, $a_{nn} = -16.4\,$fm: short-dashed line) in comparison to the $np$ 
 and $pp$ phases. For further notations, see Fig.\ref{fig:1S0200}.}
\label{fig:nn}
\end{center}
\end{figure}

\begin{figure}[htb]
\begin{center}
%\hspace{-1cm}
\begin{turn}{270}
\epsfig{file=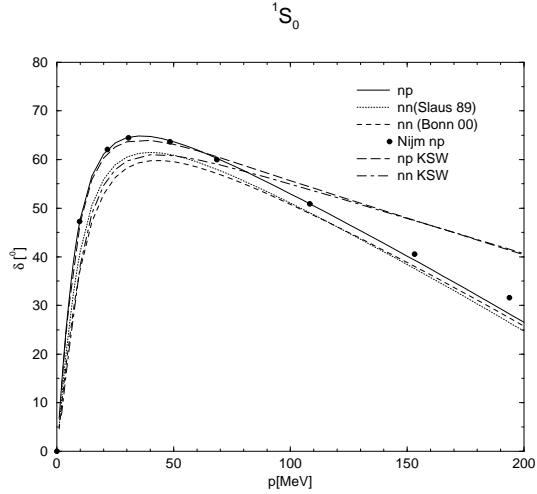, height = 7cm}
\end{turn}

\caption{Range fit for the $nn$ and $np$  $^1S_0$ phase shifts.
 The long-dashed and long-short-dashed lines refer to the $np$ and $nn$ results at 
 NLO in the KSW counting~\protect\cite{EM}. The results of the present
 approach are: $np$ solid line, $nn$ with $a_{nn} =-18.9,-16.4\,$fm:
 dotted and and short-dashed line, respectively. The filled circles
 are the Nijmegen PSA values.}
\label{fig:KSW}
\end{center}
\end{figure}

To discuss charge symmetry breaking, one has to know the $nn$
scattering length. With the ``standard value'' of $a_{nn} = -18.9 \pm
0.4\,$fm one obtains the value given in eq.(\ref{CSBval}). However,
a recent measurement performed at Bonn University using $nd$ breakup 
combined with exact solutions of the three--body problem gives a
sizeably lower value, $a_{nn} = -16.4 \pm 0.4\,$fm~\footnote{There is
a mild scatter on this result depending on the energy and
normalization which is, however, within the error bar given.}. For the
central value of the Bonn result, we get 
\beq\label{CSBBonn}
a_{\rm CSB} = -0.9~{\rm fm}~,
\eeq
which is smaller in size than the value of eq.(\ref{CSBval}) and of
{\em opposite} sign. As discussed before, we can not predict this scattering
length due to the presence of a four--nucleon LEC but rather use the
value of $a_{nn}$ as input to predict the $nn$ $^1S_0$ phase
shift shown in Fig.\ref{fig:nn} in comparison to the $np$ and $pp$
phases. Note also that CSB dies out much faster than CIB, at $E_{\rm
  lab} = 25\,$MeV the two $nn$ curves are identical. 
It is also interesting to investigate the origin
of CSB in the EFT. At NLO, there are two effects, namely the nucleon mass
difference and the short--distance physics encoded in the leading
order CSB four--nucleon interaction. While in the case based on 
eq.(\ref{CSBval}), the nucleon mass difference effect (0.33~fm) and the
contact term (1.27~fm) go in the same direction, for a scattering
length as in eq.(\ref{CSBBonn}), the short distance physics encoded
in the contact interaction (-1.21~fm) has to overcome the positive
contribution from the nucleon mass difference (0.29~fm). This becomes
more transparent if we work out the LECs as defined in
eqs.(\ref{CCSB},\ref{CCIB}) at leading order. Using the standard value
for $a_{nn}$, we have (in units of $10^4$~GeV$^{-2}$), 
\beq
C_{\rm CIB} = 0.0026~, \quad  C_{\rm CSB} = 0.0012~,
\eeq
whereas the value of the scattering length from the Bonn measurement
gives
\beq
C_{\rm CIB} = 0.0036~, \quad  C_{\rm CSB} = -0.0008~.
\eeq
We remark that the importance of the contact term contributions
to CSB is in agreement with findings in the CD--Bonn potential, where
CSB is largely driven by TPE with non--nucleonic intermediate states.
Such contributions are subsumed in the contact interactions of our approach.
It is also interesting to compare
these NLO results with ones obtained in the KSW scheme~\cite{EM}, cf.
Fig.\ref{fig:KSW}. As it has already been found in the isospin
symmetric calculations, both schemes give similar results in the low energy
region but the KSW scheme can not be applied systematically
for momenta larger than the pion mass.

%%%%%%%%%%%%%%%%%%%%%%%%%%%%%%%%%%%%%%%%%%%%%%%%%%%%%%%%%%%%%%%%%%%%%%%%%%%%%%%%%
\section{Summary}
\def\theequation{\arabic{section}.\arabic{equation}}
\setcounter{equation}{0}
\label{sec:summ}

In this paper, we have calculated charge independence and charge
symmetry breaking in the two--nucleon system based on a chiral effective 
field theory. This naturally extends the results for the $np$ system
presented in refs.\cite{EGMI,EGMII}. The results of this
investigation can be summarized as follows:

\begin{enumerate}
\item[1)] Based on a modified Weinberg power counting (as explained
  in ref.\cite{EGMI}), we have systematically included strong and
  electromagnetic isospin violation in a chiral two--nucleon potential at 
  NLO. Strong isospin violation is due to the light quark mass difference
  and can easily be incorporated by means of an external scalar source. The
  electromagnetic effects due to hard photons are represented by local
  contact interactions. Soft photons appear in loop diagrams and also generate
  the long--range Coulomb potential. The corresponding classification scheme
  for the various contributions to CIB and CSB based on the extended
  power counting at $\LOI$ and N$\LOI$ is given in section~\ref{sec:class}.
\item[2)] The resulting potential consists of two distinct pieces, the 
  strong (nuclear) potential including isospin violating effects and the Coulomb
  potential. The nuclear potential consists of one-- and two--pion exchange
  graphs (with different pion and nucleon masses), $\pi\gamma$ exchange
  diagrams and a set of  local
  four--nucleon operators (some of which are isospin symmetric, some depend
  on the quark charges and some on the quark mass difference). 
  Since the nuclear effective potential is naturally formulated in momentum space,
  we use the matching procedure developed in ref.\cite{VP} to incorporate
  the correct asymptotical Coulomb states. The necessary regularization of 
  the potential is performed on the level of the Lippmann--Schwinger equation, 
  using a sharp momentum space cut--off $\Lambda$.
\item[3)] The low--energy constants accompanying the contact interactions
  and the cut--off $\Lambda$ are determined by a {\em simultaneous}
  best fit to the S-- and P--waves of the Nijmegen phase shift analysis 
  in the $np$ and the $pp$ systems for
  laboratory energies below 50~MeV. This allows to predict these partial
  waves at higher energies and all higher partial waves.
  Most physical observables come out independent of the cut--off for $\Lambda$
  between 300 and 500~MeV. The upper limit on this range is determined by the
  $pp$ interactions. The resulting phase shifts are shown in
  figs.\ref{fig:1S0200}-\ref{fig:DFG200}, and the CIB phases
  $\delta_{\rm CIB} = \delta_{pp} - \delta_{np}$ in
  figs.\ref{fig:1S0CIB}-\ref{fig:DFGCIB}, in comparison to the
  outcome of the Nijmegen PSA and recent results based on
  the CD--Bonn potential~\cite{CD-BonnII}.
\item[4)] We have studied in detail the range expansion for the $np$ and
 the $pp$ system. For the range of cut--offs, the $pp$ scattering length
 varies modestly with $\Lambda$  due to the scheme--dependent separation of the nuclear
 and the Coulomb part, $|\delta a_{pp} / a_{pp}| \simeq 0.6~{\rm fm}/
 17~{\rm fm} \simeq 3\%$. We have dissected the various contributions
 to CIB in the scattering length, reconfirming the importance of the
 pion mass difference in the OPE. The TPE contribution is smaller in size
 but of opposite sign, and also strongly cut--off dependent.
\item[5)] We have also studied the $nn$ system, performing effective
  range fits based on the ``standard'' value for $a_{nn} = -18.9\,$fm and
  using the most recent result from deuteron break-up, 
  $a_{nn} = -16.4\,$fm~\cite{Witsch}. This leads to rather different
  physics underlying charge symmetry breaking, as discussed in
  section~\ref{sec:nn}. 
\end{enumerate}
We have shown that charge independence and charge symmetry breaking
can be systematically studied in an effective field theory based on a
modified Weinberg power counting. In the future, it would be of
interest to also study the effects of two--pion exchange diagrams
with the inclusion of explicit delta degrees of freedom to further
dissect the physics underlying CIB and CSB. In addition, one should
also include vacuum polarisation and magnetic moment corrections to
the Coulomb potential if one wants to achieve the same precision as
in the so--called modern potentials. Work along such lines is in progress.

\bigskip

%%%%%%%%%%%%%%%%%%%%%%%%%%%%%%%%%%%%%%%%%%%%%%%%%%%%%%%%%%%%%%%%%%%%%%%%%%%%%%%%%
\section*{Acknowledgements}

We are grateful to
Charlotte Elster, Walter Gl\"ockle, Johann Haidenbauer,
Ruprecht Machleidt and Bira van Kolck for useful comments and suggestions.
UGM thanks the Institute of Nuclear Theory at the University of
Washington for its hospitality and the Department of Energy for
partial support during the completion of this work.

\bigskip

%%%%%%%%%%%%%%%%%%%%%%%%%%%%%%%%%%%%%%%%%%%%%%%%%%%%%%%
\appendix
\def\theequation{\Alph{section}.\arabic{equation}}
\setcounter{equation}{0}
\section{Closer look at the two--pion exchange potential}
\label{app:ren}

In this appendix we discuss in more detail the CIB contributions
from the TPEP. In fact, we will show that one generates some
polynomial pieces at N$\LOI$~ which lead to finite shifts in
some LECs. However, the effect of these shifts will turn out 
to be numerically very small. Nevertheless,
when improving the accuracy of our potential, such effects have
to be taken into account.

\begin{figure}[htb]
\begin{center}
\begin{turn}{270}
\epsfig{file=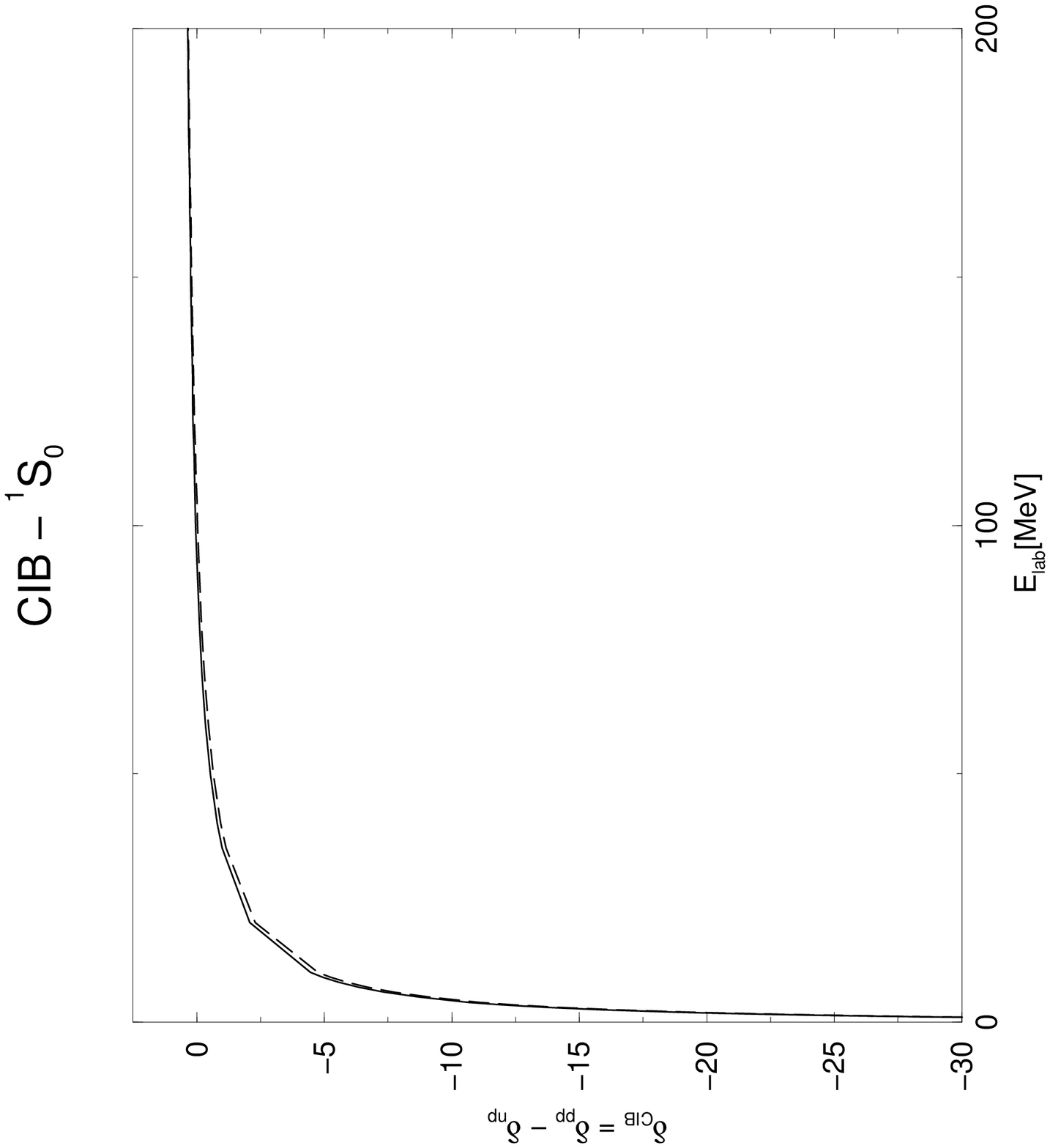, height = 7.5cm}
\end{turn}
\vspace{.5cm}
\begin{turn}{270}
\epsfig{file=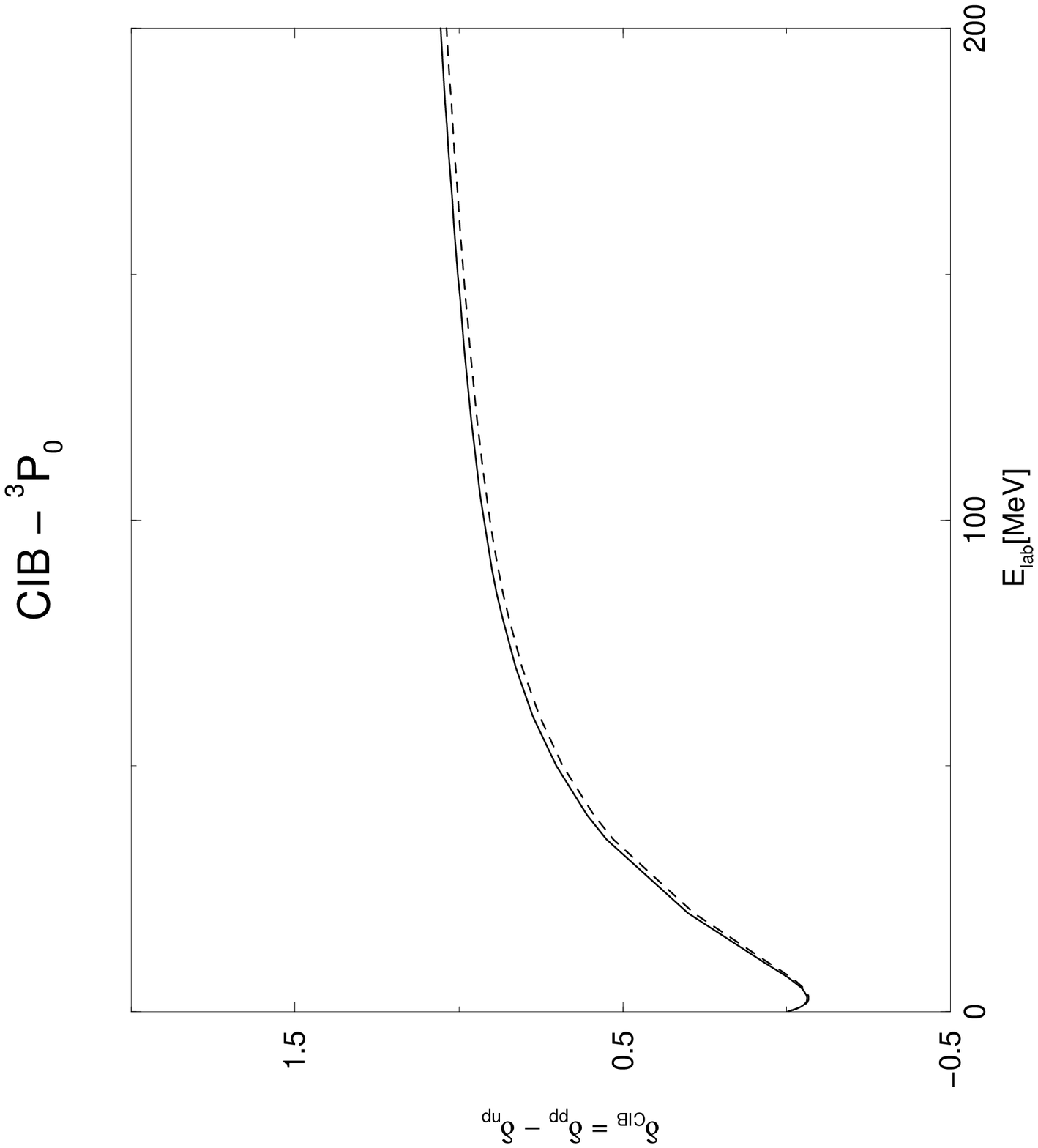, height = 7.5cm}
\end{turn}
\hspace{.5cm}
\begin{turn}{270}
\epsfig{file=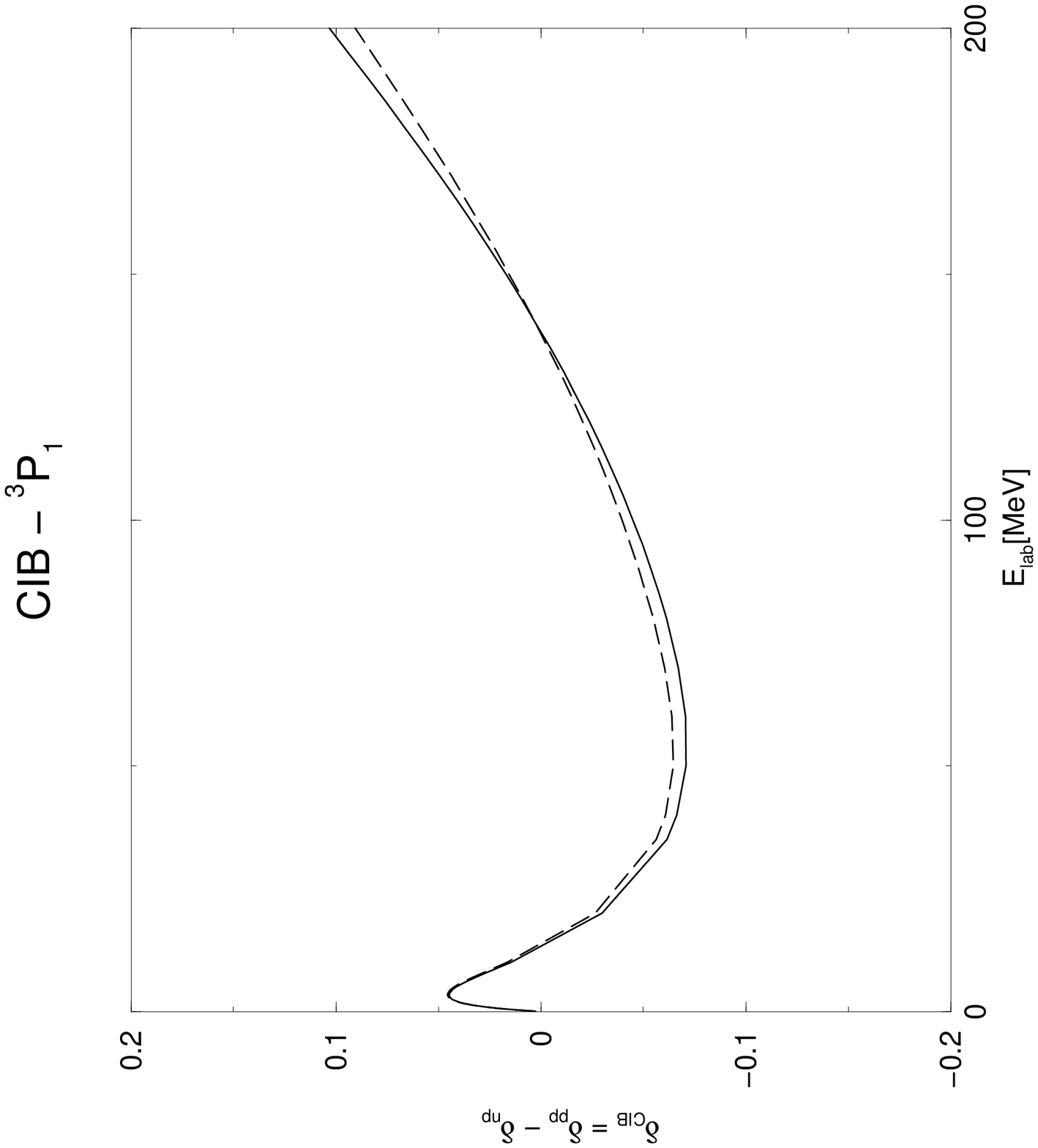, height = 7.5cm}
\end{turn}
\vspace{.5cm}
\caption{Effect of the polynomial TPEP shifts on the $^1S_0$, $^3P_0$, $^3P_1$
  waves as discussed in the appendix. The (solid)  dashed lines correspond
  to (not) incorporating  the finite shifts as given in eq.(\ref{Shift}).
}
\label{fig:shift}
\end{center}
%\vspace{-0.5cm}    
\end{figure}

\medskip \noindent
In the isospin symmetric case, we can write the NLO TPEP 
$V^{(2)}_{2\pi, {\rm 1-loop}}$in the form, see also eq.(\ref{TPEP}),
\beq\label{V2p}
V^{(2)}_{2\pi, {\rm 1-loop}} = V^{\rm TPEP}_{\rm non-polynom} + (S_1 + S_2 \,q^2) \, (\vec
  \tau_1 \cdot  \vec \tau_2 ) + S_3 \, \biggl[ ( \vec \sigma_1 \cdot \vec q \, ) \,
 ( \vec \sigma_2 \cdot \vec q \, ) - ( \vec \sigma_1 \cdot\vec
 \sigma_2 ) \, q^2 \biggr]~,
\eeq 
with
\beqa 
S_1 &=& \frac{1}{384 \pi^2 F_\pi^4} \biggl\{ -18\Mpz (5g_A^4 -2g_A^2)
\ln \frac{\Mp}{\varepsilon} - \Mpz (61g_A^4-14g_A^2+4) \no\\
&& \qquad\qquad\qquad \qquad\qquad\qquad + 18\Mpz (5g_A^4 -2g_A^2) J_0 - 3(3g_A^4
-2g_A^2) J_2 \biggr\}~,\\
S_2 &=& \frac{1}{384 \pi^2 F_\pi^4} \biggl\{ (-23g_A^4 +10g_A^2 +1)
\ln \frac{\Mp}{\varepsilon} - {1\over 2} (13g_A^4 + 2g_A^2) 
+ (23g_A^4 -10g_A^2-1) J_0 \biggr\},\\
S_3 &=& -\frac{3g_A^4}{64 \pi^2 F_\pi^4} \biggl\{ \ln
\frac{\Mp}{\varepsilon} + \frac{1}{3} - J_0 \biggr\}~,
\eeqa
in terms of the divergent loop functions $J_{0,2}$, the IR regulator
$\varepsilon$ (for details, see ref.\cite{EGMII}) and the non--polynomial part is given in
eq.(\ref{TPEP}). From here on, we concentrate on the polynomial terms,
which have the genuine structure
\beq
V^{\rm TPEP}_{\rm polynom} = V^{0}_{\rm polynom} + V^{1}_{\rm polynom} \,
\vec{\tau}_1 \cdot \vec{\tau}_2~.
\eeq
Again, the pion mass difference $\Delta M_\pi^2 \sim \alpha$ will lead
to CIB. Following the same arguments as used in section~\ref{sec:pot},
up to corrections of order $\alpha^2$, one uses the various pion masses
in the polynomial pieces as explained for the non--polynomial ones. 
To be specific, consider the $T=1$  $np$ case. We denote the
corresponding LECs by $C_S, C_T, C_i$. Putting in the appropriate pion
masses, one finds for the isovector piece
\beqa
V^{1}_{{\rm polynom},pp,nn} - V^{1}_{{\rm polynom},np} &=&  
V^{1}_{\rm polynom} (M_{\pi^\pm}) - V^{1}_{\rm polynom} (M_{\pi^0})
\nonumber \\
&=& (A + B\, q^2) \, \vec{\tau}_1 \cdot \vec{\tau}_2~.
\eeqa
The term $\sim A$ leads to a relative shift between $C_{pp}^{\rm LO}$,
$C_{nn}^{\rm LO}$, and $C_{np}^{\rm LO}$ in the $^1S_0$ wave. However,
we do not need to know the precise value of this shift (and thus of the
constant $A$) since these LECs are any way fitted independently. To
the contrary, one has to know the value of $B$,
\beq
B = {1 \over 384\pi^2 F_\pi^4} \biggl(-23g_A^4 + 10g_A^2 +1 \biggr) \,
\ln \frac{M_{\pi^\pm}}{M_{\pi^0}} = -4.98~{\rm GeV}^{-4}~.
\eeq
This leads to finite shifts in the NLO LECs,
\beq
\Delta C_{^sL_j}^{\rm NLO} = C_{pp,nn,^sL_j}^{\rm NLO}
- C_{np,^sL_j}^{\rm NLO}~,
\eeq
in the usual spectroscopic notation. We find
\beq
\Delta C_{^1S_0}^{\rm NLO} = 4 \pi B~, \quad 
\Delta C_{^3P_j}^{\rm NLO} = - \frac{8 \pi}{3} B~, 
\eeq
so that
\beqa
C_{pp,nn,^1S_0}^{\rm NLO} = \biggl( C_{np,^1S_0}^{\rm NLO}
-0.0063\biggr) \cdot 10^4 \, {\rm GeV}^{-4}~, \nonumber \\
C_{pp,nn,^3P_j}^{\rm NLO} = \biggl( C_{np,^3P_j}^{\rm NLO}
+0.0042\biggr) \cdot 10^4 \, {\rm GeV}^{-4}~.\label{Shift}
\eeqa
The effect of these shifts is numerically very small, as demonstrated
for the $^1S_0$, $^3P_0$ and $^3P_1$ waves in fig.\ref{fig:shift}.

\newpage

%\newpage
%%%%%%%%%%%%%%%%%%%%%%%%%%%%%%%%%%%%%%%%%%%%%%%%%%%%%%%%%%%%%%%%%%%%%%%%%%%%%%%

%%\end{document}

\newpage

\section*{Figures}

$\,$

\vspace{1cm}

\vspace{3cm}

\vspace{-1.5cm}    

\newpage

\newpage

\end{document}